\documentclass[manuscript,natbib]{aastex}

\usepackage[]{natbib}

\newcommand{\vk}{(V-K$_s$)}

\newcommand{\vrad}{$v_R$}

\newcommand{\logl}{log(L/L$_{\odot}$)}

\newcommand{\loglxlbol}{log($L_{\rm X}$/$L_{bol}$)}
\newcommand{\teff}{T$_{\rm eff}$}

\newcommand{\pmra}{$\mu_{\alpha}$}
\newcommand{\pmdec}{$\mu_{\delta}$}
\newcommand{\kms}{km\,s$^{-1}$}
\newcommand{\masyr}{mas\,yr$^{-1}$}

\newcommand{\mv}{M$_V$}

\newcommand{\msun}{M$_{\odot}$}
\newcommand{\rsun}{R$_{\odot}$}

\newcommand{\ergs}{erg\,s$^{-1}$}

\newcommand{\Mbol}{M$_{\rm bol}$}
\newcommand{\mbol}{m$_{\rm bol}$}
\newcommand{\vsini}{$v$sin$i$}

\slugcomment{Accepted to Astronomical Journal}
\shorttitle{Fomalhaut C}
\shortauthors{Mamajek et al.}
\begin{document}

\title{The Solar Neighborhood XXX. Fomalhaut C}

\author{Eric E. Mamajek\altaffilmark{1,2}, Jennifer
  L. Bartlett\altaffilmark{3}, Andreas Seifahrt\altaffilmark{4}, Todd
  J. Henry\altaffilmark{5}, Sergio B. Dieterich\altaffilmark{5}, John
  C. Lurie\altaffilmark{5}, Matthew A. Kenworthy\altaffilmark{6},
  Wei-Chun Jao\altaffilmark{5}, Adric R. Riedel\altaffilmark{7,8},
  John P. Subasavage\altaffilmark{9}, Jennifer
  G. Winters\altaffilmark{5}, Charlie T. Finch\altaffilmark{3}, Philip
  A. Ianna\altaffilmark{10}, Jacob Bean\altaffilmark{4}}
\altaffiltext{1}{Department of Physics and Astronomy, 
University of Rochester, Rochester, NY 14627, USA} 
\altaffiltext{2}{Cerro Tololo Inter-American Observatory,
Casilla 603, La Serena, Chile}
\altaffiltext{3}{US Naval Observatory, 3450 Massachusetts Ave., NW, Washington, DC, 20392, USA}
\altaffiltext{4}{Department of Astronomy and Astrophysics, 
University of Chicago, IL, 60637, USA}
\altaffiltext{5}{Department of Physics and Astronomy, Georgia State University, Atlanta, GA 30302-4106, USA}
\altaffiltext{6}{Leiden Observatory, Leiden University, P.O. Box 9513, 2300 RA Leiden, The Netherlands}
\altaffiltext{7}{Department of Physics \& Astronomy, Hunter College, 695 Park Avenue, New York, NY 10065, USA}
\altaffiltext{8}{Department of Astrophysics, American Museum of Natural History, Central Park West at 79th Street, New York, NY 10034, USA}
\altaffiltext{9}{US Naval Observatory, Flagstaff Station, P.O. Box 1149, Flagstaff, AZ 86002-1149, USA}
\altaffiltext{10}{Department of Astronomy, University of Virginia, PO Box 400325, Charlottesville, VA 22904-4325, USA}

\email{emamajek@pas.rochester.edu}

\begin{abstract} 
  LP 876-10 is a nearby active M4 dwarf in Aquarius at a distance of
  7.6 pc. The star is a new addition to the 10-pc census, with a
  parallax measured via the Research Consortium on Nearby Stars
  (RECONS) astrometric survey on the Small \& Moderate Aperture
  Research Telescope System's (SMARTS) 0.9-m telescope.  We
  demonstrate that the astrometry, radial velocity, and photometric
  data for LP 876-10 are consistent with the star being a third,
  bound, stellar component to the Fomalhaut multiple system, despite
  the star lying nearly 6$^{\circ}$ away from Fomalhaut A in the sky.
  The 3D separation of LP 876-10 from Fomalhaut is only
  0.77\,$\pm$\,0.01 pc, and 0.987\,$\pm$\,0.006 pc from TW PsA
  (Fomalhaut B), well within the estimated tidal radius of the
  Fomalhaut system (1.9 pc). LP 876-10 shares the motion of Fomalhaut
  within $\sim$1 km/s, and we estimate an interloper probability of
  $\sim$10$^{-5}$.  Neither our echelle spectroscopy nor astrometry
  are able to confirm the close companion to LP 876-10 reported in the
  Washington Double Star Catalog (WSI 138). We argue that the Castor
  Moving Group to which the Fomalhaut system purportedly belongs, is
  likely to be a dynamical stream, and hence membership to the group
  does not provide useful age constraints for group members. LP 876-10
  (Fomalhaut C) has now risen from obscurity to become a rare example
  of a field M dwarf with well-constrained age (440\,$\pm$\,40 Myr)
  and metallicity. Besides harboring a debris disk system and
  candidate planet, Fomalhaut now has two of the widest known stellar
  companions.
\end{abstract}

\keywords{
binaries: visual ---
Stars: activity ---
Stars: fundamental parameters --- 
Stars: individual (LP 876-10, Fomalhaut, TW PsA) ---
Stars: rotation
}
\section{Introduction}

Fomalhaut is an important nearby A3~V star, containing a large
resolved dusty debris disk \citep{Gillett86, Kalas05} and a candidate
extrasolar planet \citep{Kalas08, Kalas13, Quillen06}.  Fomalhaut has
previously had at least two stars suggested to be companions.
\citet{See1898} reported a 14th magnitude stellar companion to
Fomalhaut at 30'' separation, however this star was later deemed a
background star by \citet{Burnham78}
\footnote{\citet{See1898} reported a single observation of a 14th
  magnitude companion at $\theta$ = 36$^{\circ}$.2, separation
  29''.98, at epoch 1896.706. Dubbed ``$\lambda_1$ 478'' by See, this
  object appears to have largely disappeared from the literature, and
  does not appear in the modern Washington Double Star catalog.  The
  only subsequent mentions of this companion that we found are in the
  \citet{Burnham06} compendium of double stars (Entry \#12071 is
  listed as "See 478"), and in two popular books
  \citep{Allen63,Burnham78}.  \citet{Burnham78} stated ``{\it it
    appears to be merely a faint field star, having no real connection
    with Fomalhaut}.''  Based on the \citet{vanLeeuwen07} Hipparcos
  astrometry for Fomalhaut A, we estimate that Fomalhaut A has moved
  35'' since See's observation, and was at ICRS position 22:57:36.44
  -29:37:03.0 at epoch 1896.706. See's reported position angle and
  offset corresponds to $\Delta$$\alpha$ = +17".7, $\Delta$$\delta$ =
  +24".2, hence if this object were stationary, we would predict its
  ICRS position to be near 22:57:37.8 -29:36:39. No catalogued object
  appears near this position.  Examination of Fig. 3 of
  \citet{Marengo09}, an IRAC 4.5\,$\mu$m full-array, roll-subtracted
  image taken with Spitzer Space Telescope, shows no obvious point
  source either at the position See reported, nor where See's star
  would appear if it were comoving with Fomalhaut.  Given that 1) See
  only reported a single observation, 2) no subsequent literature
  characterized the object, and 3) we were unable to find the star in
  the Spitzer imagery and other modern catalogs, we conclude that
  See's reported companion to Fomalhaut was likely spurious.}.
\citet{Luyten38} reported discovery of a K-type common proper motion
companion to Fomalhaut: TW PsA (HR 8721). The physicality of the
Fomalhaut-TW PsA binary system was investigated by \citet{Barrado97}
and \citet{Mamajek12}, and both studies concluded that the pair
comprise a physical binary.  \citet{Mamajek12} estimated that
Fomalhaut and TW PsA have a true separation of only 0.28 pc and share
velocities within 0.1\,$\pm$\,0.5 \kms, consistent with constituting a
bound system.  \citet{Mamajek12} estimated the age for the Fomalhaut
binary system to be 440\,$\pm$\,40 Myr based upon multiple age
indicators, with the isochronal age of Fomalhaut A and the
gyrochronology age of Fomalhaut B providing the most weight.\\

During the preparation of the \citet{Mamajek12} article, another
  neighboring star was identified that appeared to share motion with
  Fomalhaut and TW PsA: LP 876-10 (NLTT 54872, WT 2282, 2MASS
  J22480446-2422075, PM I22480-2422).  LP 876-10 is a high proper
motion star first catalogued as such by \citet{Luyten80}, situated
5$^{\circ}$.67 NW (20407\farcs6; PA = 337$^{\circ}$.91) of
Fomalhaut. At the time of writing \citet{Mamajek12} there was
  insufficient evidence to test whether LP 876-10 was truly associated
  with the Fomalhaut binary, with the main evidence being the
  coincidental proper motion and photometric distance.  In this
  contribution, we combine newly determined accurate astrometric and
  radial velocity measurements for LP 876-10 to demonstrate that it
  too, like TW PsA, appears to be a distant companion of Fomalhaut,
  and should be considered ``Fomalhaut C''.\\

\section{Analysis\label{sec:data}}

\begin{figure}[htb!]
\epsscale{0.3}
\plotone{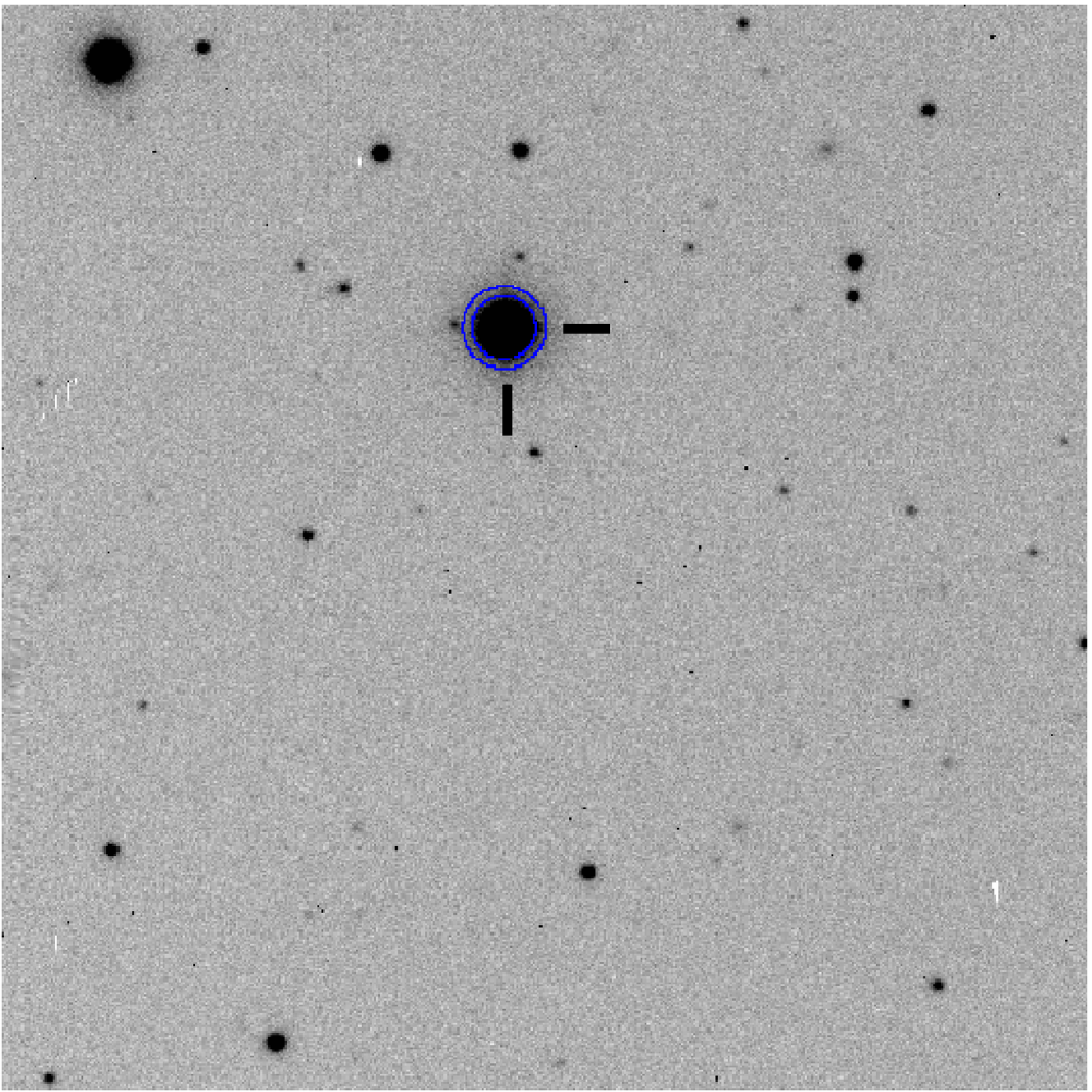}
\plotone{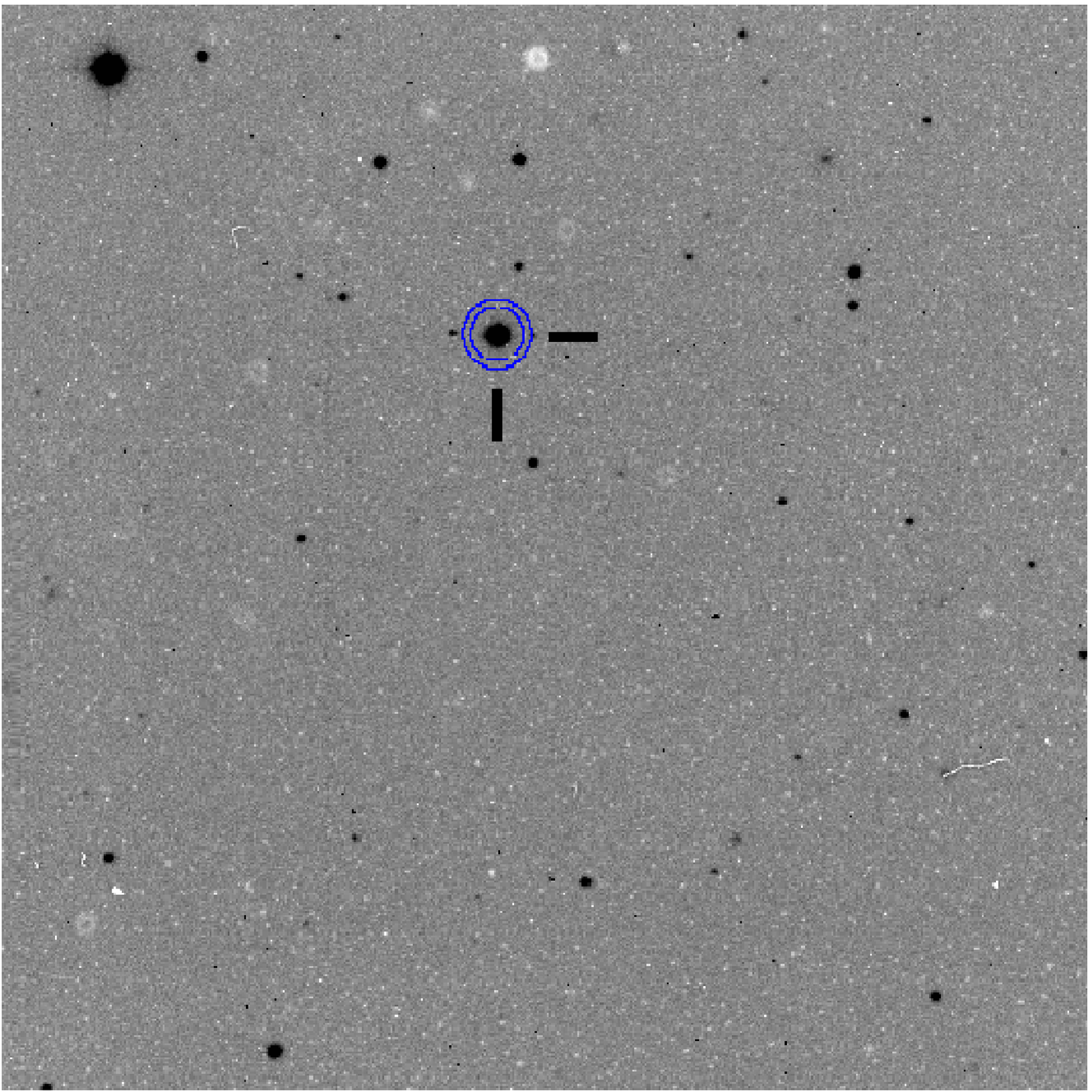}
\caption{{\it Left:} Kron-Cousins I-band image of LP 876-10 taken with
  SMARTS 0.9-m telescope in 2004. {\it Right:} Johnson V-band image of
  LP 876-10 taken with SMARTS 0.9-m telescope in 2012.  Star is
  circled in blue and is near 22:48:04.5 -24:22:08 (J2000). Field of
  view for both images is 6.8 arcminutes square. North is up and east
  is left. Fomalhaut A and B are not in the field of view.
  \label{fig:finder}}
\end{figure}

The stellar parameters for Fomalhaut, TW PsA, and LP 876-10 are
summarized in Table \ref{tab:params}. Finder charts for LP 876-10 are
provided in Figure \ref{fig:finder}.\\


\begin{deluxetable}{lccccc}
\tabletypesize{\scriptsize}
\setlength{\tabcolsep}{0.03in}
\tablewidth{0pt}
\tablecaption{Stellar Parameters\label{tab:params}}
\tablehead{
{(1)}   &{(2)}          & {(3)}       &{(4)}        &{(5)}    & {(6)}\\
{Value} & {$\alpha$ PsA}&{TW PsA}     &{LP 876-10}  & {Units} & {Ref.}\\
{...}   &{Fomalhaut A}  &{Fomalhaut B}&{Fomalhaut C}&{...}
}
\startdata
$\alpha_{ICRS}$(J2000)  &  344.411773            &  344.099277              &  342.018632            &  deg          &  1,1,2\\
$\delta_{ICRS}$(J2000)  &  -29.621837            &  -31.565179              &  -24.368872            &  deg          &  1,1,2\\
Parallax                &  129.81\,$\pm$\,0.47   &  131.42\,$\pm$\,0.62     &  132.07\,$\pm$\,1.19   &  mas          &  1,1,3\\
Distance                &  7.704\,$\pm$\,0.028   &  7.609\,$\pm$\,0.036     &  7.572\,$\pm$\,0.068   &  pc           &  1,1,3\\
\pmra\,                 &  329.95\,$\pm$\,0.50   &  331.11\,$\pm$\,0.65     &  333.8\,$\pm$\,0.5     &  \masyr       &  1,1,3\\
\pmdec\,                &  -164.67\,$\pm$\,0.35  &  -158.98\,$\pm$\,0.48    &  -177.5\,$\pm$\,0.7    &  \masyr       &  1,1,3\\
\vrad\,                 &  6.5\,$\pm$\,0.5       &  6.6\,$\pm$\,0.1         &  6.5\,$\pm$\,0.5       &  \kms         &  4,5,3\\
m$_V$                   &  1.155\,$\pm$\,0.005   &  6.488\,$\pm$\,0.012     &  12.62\,$\pm$\,0.01    &  mag          &  6,6,3\\
M$_V$                   &  1.72\,$\pm$\,0.01     &  7.08\,$\pm$\,0.02       &  13.21\,$\pm$\,0.02    &  mag          &  3,3,3\\
Period$_{Rot}$          &  ...                   &  10.3                    &  0.466                 &  day          &  7,3\\
SpT                     &  A3~Va                 &  K4~Ve                   &  M4~V                  &  ...          &  8,9,10\\
\teff\,                 &  8590\,$\pm$\,73       &  4594\,$\pm$\,80         &  3132\,$\pm$\,65\,K    &  K            &  11,12,3\\
$f_{bol}$               &  8.96\,$\pm$0.25       &  0.10075                 &  0.00257(5)            &  nW~m$^{-2}$  &  13,12,3\\
X$_{gal}$               &  3.06                  &  3.14                    &  3.01                  &  pc           &  11,11,3\\
Y$_{gal}$               &  1.14                  &  0.90                    &  1.86                  &  pc           &  11,11,3\\
Z$_{gal}$               &  -6.98                 &  -6.88                   &  -6.70                 &  pc           &  11,11,3\\
$\Delta_{com}$          &  0.05                  &  0.24                    &  0.77                  &  pc           &  3,3,3\\
U                       &  -5.71\,$\pm$\,0.16    &  -5.69\,$\pm$\,0.06      &  -5.34\,$\pm$\,0.19    &  \kms         &  11,11,3\\
V                       &  -8.26\,$\pm$\,0.28    &  -8.16\,$\pm$\,0.07      &  -7.58\,$\pm$\,0.28    &  \kms         &  11,11,3\\
W                       &  -11.04\,$\pm$\,0.38   &  -10.96\,$\pm$\,0.08     &  -11.85\,$\pm$\,0.39   &  \kms         &  11,11,3\\
$\Delta$S               &  0                     &  0.13\,$\pm$\,0.51       &  1.12\,$\pm$\,0.72     &  \kms         &  3,3\\
\logl                   &  1.221\,$\pm$\,0.013   &  -0.723\,$\pm$\,0.029    &  -2.337\,$\pm$\,0.010  &  dex          &  11,11,3\\
Mass                    &  1.92\,$\pm$\,0.02     &  0.73$^{+0.02}_{-0.01}$   &  0.18\,$\pm$\,0.02     &  \msun        &  11,12,3\\
$\theta$                &  0                     & 7062\farcs.7          & 20407\farcs.6              & arcsec        & 14,14\\
PA                      &  0                     & 187$^{\circ}$.88         & 337$^{\circ}$.91        & deg           & 14,14\\
\enddata
\tablecomments{$\Delta_{com}$ is the approximate 3D separation between
  the star and the system's center of mass.  $\Delta$S is the
  difference in velocity compared to Fomalhaut A. Masses are estimated
  using evolutionary tracks and are not dynamically measured. $\theta$
  is the projected angular separation from Fomalhaut, and PA is the
  position angle as measured north through east.  References:
(1) \citet{vanLeeuwen07} (distance = 1/parallax), 
(2) \citet{Roeser10} (PPMXL), 
(3) this paper, 
(4) \citet{Gontcharov06},
(5) \citet{Nordstrom04}, 
(6) \citet{Mermilliod94},
(7) \citet{Busko78},
(8) \citet{GrayGarrison89A} (standard), 
(9) \citet{Keenan89}, 
(10) \citet{Scholz05},
(11) \citet{Mamajek12}, 
(12) \citet{Casagrande11}, 
(13) \citet{Davis05},
(14) this paper, using positions from \citet{vanLeeuwen07} and \citet{Roeser10}. 
In the Table and throughout the paper, Galactic velocities $U$ and positions $X$ are 
defined towards the Galactic center, $V$ and $Y$ are in the direction of Galactic 
rotation, and $W$ and $Z$ is towards the North Galactic pole.}
\end{deluxetable}

\subsection{Photometry \label{sec:phot}}

Adopted optical and infrared magnitudes from 0.4$\mu$m (B-band) to 22
$\mu$m (W4-band) are compiled in Table \ref{tab:phot}.  V-band
magnitudes of 12.618\,$\pm$\,0.012 \citep{Reid03} and
12.62\,$\pm$\,0.02 \citep[UCAC4;][]{Zacharias13, Henden12} have been
previously reported for LP 876-10.  Additionally, we measure Johnson V
= 12.59\,$\pm$\,0.03 based on 3 photometric nights of imaging with the
Small and Moderate Aperture Research Telescope System's (SMARTS) 0.9-m
telescope taken during 2004-2006, using 14'' diameter aperture and the
standards of \citet{Landolt92,Landolt07}. This measurement is
consistent with preliminary values previously reported from this
program \citep{Bartlett07PhD, Bartlett07AAS}. \citet{Jao05} and
\citet{Winters11} describe the REsearch Consortium On Nearby Stars
(RECONS)\footnote{www.recons.org} photometry
program. \citet{Pojmanski97} presents time series V-band ASAS
photometry for this star (349 observations between UT 22 November 2000
and UT 7 October 2009 with quality flag A) with mean V = 12.62 and rms
scatter of 0.06 mag. The photometric errors of the individual
measurements are typically $\sim$0.03 mag, so approximately $\sim$0.05
mag of the scatter in the reported V magnitudes appear to be due to
intrinsic stellar variability. All of these previously mentioned V
magnitudes are calibrated to the Johnson system, either through
Landolt standards (Reid et al. 2003, APASS/UCAC4, RECONS) or Hipparcos
(ASAS). The SuperWASP project took many photometric measurements of LP
876-10 in a V-band calibrated to the Tycho-2 $V_T$ photometric system
\citep{Pollacco06, Butters10, Hog00}. These photometry are later
discussed in Sec. 2.8 for the purposes of measuring the rotation
period, but were not included in our assessment of the mean Johnson V
magnitude. Based on photometry measured independently by
\citet{Reid03}, \citet{Henden12}, \citet{Pojmanski97}, and the RECONS
observations with the SMARTS 0.9-m telescope, we adopt a mean V
magnitude of 12.62\,$\pm$\,0.01 mag.\\


\begin{deluxetable}{lcl}
\tabletypesize{\scriptsize}
\setlength{\tabcolsep}{0.03in}
\tablewidth{0pt}
\tablecaption{LP 876-10 Photometry\label{tab:phot}}
\tablehead{
{(1)}   &{(2)}  & {(3)}\\
{Band}  & {Mag} & {Ref.}}
\startdata
B     & 14.31\,$\pm$\,0.01  & 1,2\\
V     & 12.62\,$\pm$\,0.01   & 3\\
R$_{KC}$ & 11.31\,$\pm$\,0.03 & 4\\
I$_{KC}$ &  9.61\,$\pm$\,0.03 & 4\\
J     &  8.075\,$\pm$\,0.023 & 5\\
H     &  7.527\,$\pm$\,0.055 & 5\\
K$_s$ &  7.206\,$\pm$\,0.021 & 5\\
W1    &  6.911\,$\pm$\,0.034 & 6\\
W2    &  6.803\,$\pm$\,0.022 & 6\\
W3    &  6.705\,$\pm$\,0.016 & 6\\
W4    &  6.497\,$\pm$\,0.058 & 6
\enddata
\tablecomments{References: (1) \citet{Reid03}, (2) APASS photometry
  \citep{Henden12} reported in UCAC4 catalog \citet{Zacharias13}
  UCAC4, (3) mean of photometry from \citet{Reid03}, \citet{Henden12},
  \citet{Pojmanski97}, and measured in this study using SMARTS 0.9-m
  telescope; see discussion in \S\ref{sec:phot}, (4) this paper, (5)
  \citet{Skrutskie06} (2MASS PSC), (6) \citet{Wright10} (WISE).
    The Johnson $B$ from APASS \citep{Henden12}, and the Johnson $V$
    and Kron-Cousins $R_{KC}I_{KC}$ photometry from RECONS
    \citep{Jao05,Winters11}, are all photometrically calibrated to
    \citet{Landolt92} standard stars.}
\end{deluxetable}

\subsection{Parallax and Proper Motion \label{sec:plx}}

The parallax and proper motion of LP 876-10 have been measured during
the long-term astrometry program carried out by RECONS at the SMARTS
0.9-m telescope. \citet{Jao05} describes the astrometry program,
however we briefly summarize the program here.  A filter is selected
from the Johnson-Kron-Cousins $V$$R_{RC}$$I_{RC}$ filterset that
provides a well-exposed reference field that, ideally, encircles the
target star.  Throughout the course of the observations, the same
pointing (to within a few pixels) and filter are used.  Centroids for
the reference field and parallax star are extracted using SExtractor
\citep{Bertin96} and corrected for differential color refraction using
$V$$R_{RC}$$I_{RC}$ photometry of the reference and science target
stars (see Section 2.1).  Relative parallax and proper motion of the
target star are solved for using the Gaussfit
program\footnote{Available from the HST Astrometry Team at
  ftp://clyde.as.utexas.edu/pub/gaussfit/.}.  Correction from relative
to absolute parallax is done by estimating the mean distance to the
reference field stars, again, using $V$$R_{RC}$$I_{RC}$ photometry and
the photometric distance relations of \citet{Henry04}.

\begin{figure}[htb!]
\epsscale{1.0}
\plotone{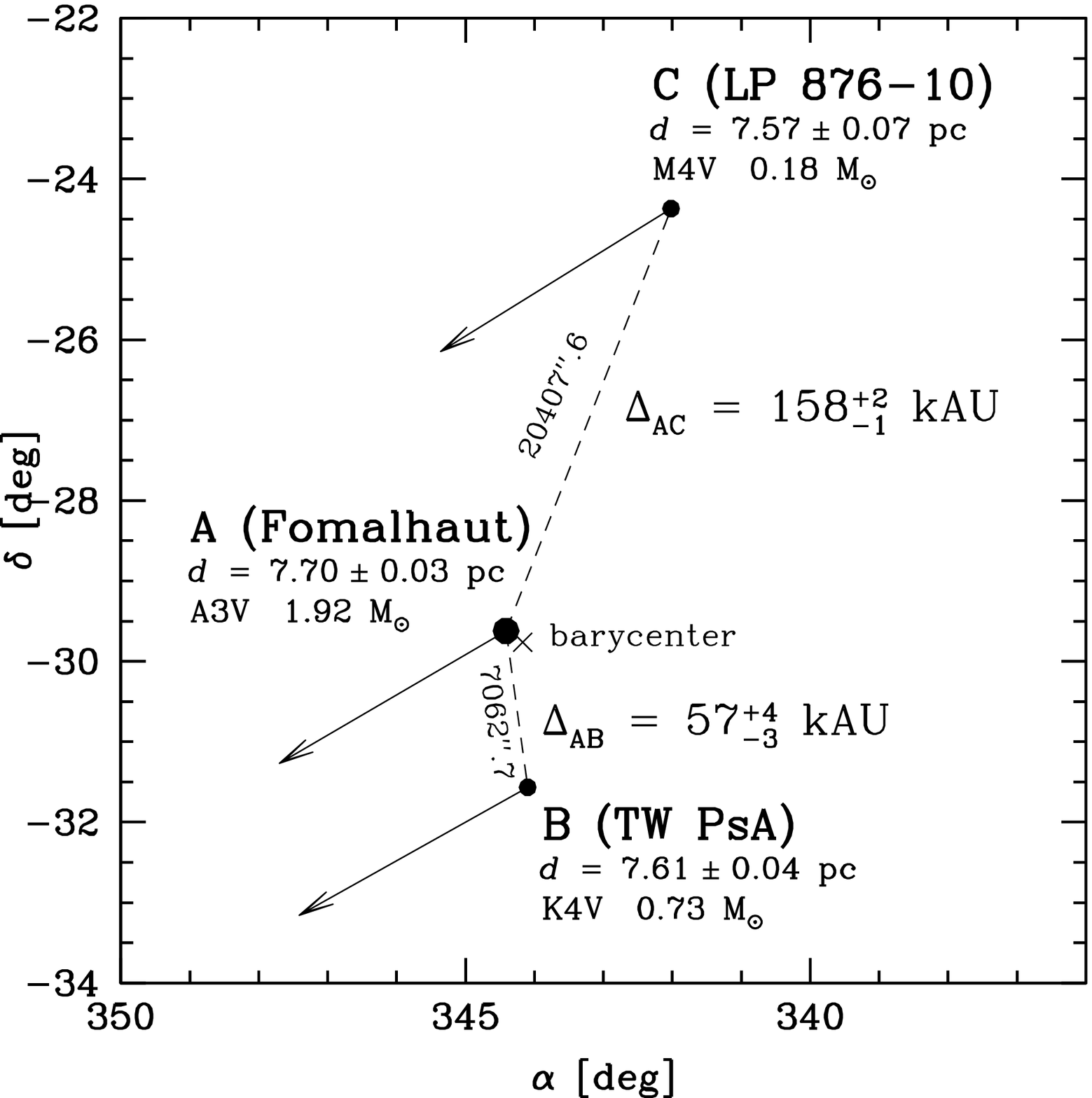}
\caption{Positions and proper motion vectors for Fomalhaut A, B (TW
  PsA), and C (LP 876-10). The system barycenter is estimated to be at
  the position marked with an X, at distance 7.67 pc (more details are
  discussed in \S2.2). The 3D separations ($\Delta$) between A-B and
  A-C are listed in thousands of AU. The positions, parallaxes,
  distances, and proper motions are compiled in Table 1.
\label{fig:xy}}
\end{figure}

LP 876-10 was included in the RECONS astrometric survey due to its
close predicted photometric distance \citep[7.2\,$\pm$\,0.8 pc;
][]{Reid03}, which is consistent with preliminary parallax solutions
from this program \citep{Bartlett07PhD, Bartlett07AAS}.  Based on 25
astrometric nights from 2004 to 2012, we derive an absolute
trigonometric parallax of 132.07\,$\pm$\,1.19 mas and a proper motion
of 378.1\,$\pm$\,0.4 \masyr\, at position angle (PA)
118$^{\circ}$.0\,$\pm$\,0$^{\circ}$.1 east of north.  When the proper
and parallactic motions are removed from the star's position, the
residuals show no hint of curvature or any pattern that would suggest
the existence of an unseen companion (see \S2.4).  At distance $d$ =
7.57\,$\pm$\,0.07 pc, the 3D separation of LP 876-10 from Fomalhaut is
only 0.77\,$\pm$\,0.01 pc (158$^{+2}_{-1}$ kAU), and from TW PsA it
lies only 0.987$^{+0.006}_{-0.005}$ pc (203\,$\pm$\,1 kAU) away
\citep[Fomalhaut B;][]{Mamajek12}.  Figure \ref{fig:xy} summarizes the
positions, separations, and proper motion vectors for Fomalhaut, TW
PsA, and LP 876-10.\\


Blinking images suggests that the neighboring high proper motion star
LP 876-11 could be a proper motion companion to LP 876-10; LP 876-11
is located 1'.8 away from LP 876-10 at 42$^{\circ}$ east of north.
However, we determine a proper motion for LP 876-11 of
321.3\,$\pm$\,0.7 \masyr\, at PA 143$^{\circ}$.0\,$\pm$\,0$^{\circ}$.2
east of north, which is inconsistent with the measured motion for LP
876-10. Using twelve color-magnitude relations from \citet{Henry04},
we estimate a photometric distance to LP 876-11 of 730\,$\pm$\,120
pc. We measure a trigonometric parallax of LP 876-11
\footnote{Photometry for LP 876-11 from 2 nights of observations: $V$
  = 17.72, $R_{KC}$ = 16.90, and $I_{KC}$ = 16.04.} of 1\,$\pm$\,2
mas, consistent with the photometric distance. We conclude that LP
876-11 is not physically associated with LP 876-10.\\


\begin{deluxetable}{lll}
\tabletypesize{\scriptsize}
\setlength{\tabcolsep}{0.03in}
\tablewidth{0pt}
\tablecaption{Proper Motions for LP 876-10\label{tab:pm}}
\tablehead{
{(1)}           & {(2)}    & {(3)}\\
{Reference}     & {\pmra}  & {\pmdec}\\
{...}           & {\masyr} & {\masyr}}
\startdata
\citet{Wroblewski99}           & 290\,$\pm$\,9       & -176\,$\pm$\,6\\
\citet{Hambly01} (SuperCOSMOS, UKST blue) & 326.9\,$\pm$\,15.66 & -191.9\,$\pm$\,19.21\\
\citet{Hambly01} (SuperCOSMOS, UKST red) & 329.9\,$\pm$\,13.12 & -187.6\,$\pm$\,14.10\\
\citet{Hambly01} (SuperCOSMOS, UKST IR) & 344.2\,$\pm$\,27.63 & -184.0\,$\pm$\,16.00\\
\citet{Monet03} (USNO-B1.0)    & 312\,$\pm$\,2       & -180\,$\pm$\,3\\
\citet{Salim03}                & 323.3\,$\pm$\,5.5   & -174.8\,$\pm$\,5.5\\
\citet{Roeser08} (PPMX)        & 311.2\,$\pm$\,11.6  & -181.4\,$\pm$\,14.0\\
\citet{Roeser10} (PPMXL)       & 322.6\,$\pm$\,4.8   & -183.9\,$\pm$\,4.8\\
\citet{Lepine11}               & 325\,$\pm$\,8       & -181\,$\pm$\,8\\
\citet{Zacharias13} (UCAC4)    & 323.0\,$\pm$\,8     & -174.8\,$\pm$\,8\\
this paper                     & 333.84\,$\pm$\,0.51 & -177.51\,$\pm$\,0.51
\enddata
\end{deluxetable}

\subsection{Radial Velocity\label{sec:RV}}

A spectrum of LP~876-10 was taken with the CRIRES spectrograph on the
8.4-m VLT UT1 (Antu) telescope on UT date 16 June 2009 as part of a
near-infrared radial velocity survey of nearby late-type M dwarfs
\citep{Bean10}.  The CRIRES spectrum has wavelength coverage
2.292--2.349 $\mu$m over the effective 4096 x 512 focal plane
detector, a mosaic of four Aladdin III InSb arrays \citep{Kaufl04}.
The slit width was 0\arcsec.2, yielding a resolving power of R $\simeq$
  100,000 (resolution is 3 \kms\, at 2 pixel sampling). The signal to
  noise ratio in the continuum of the spectrum was $\sim$170-220.  By
fitting a broadened and shifted PHOENIX model spectrum from the GAIA
V2.0 library \citep{Hauschildt99, Brott05} to the spectrum of
LP~876-10, we determine a sizeable projected rotation velocity of
\vsini\, = 22\,$\pm$\,2 \kms; a heliocentric radial velocity of
+6.5\,$\pm$\,0.5 \kms\ was also measured.  Slit viewer images of~LP
876-10 appear point-like, and there is no sign of duplicity in the
CRIRES spectrum. A more detailed spectroscopic analysis of LP
  876-10 will be presented in a forthcoming paper (Seifahrt et al., in prep.).\\

\subsection{Duplicity \label{sec:binarity}}

While neither the astrometry nor the spectroscopy data are consistent
with LP 876-10 being a binary, it is listed as a double star in the
Washington Double Star catalog \citep[WDS;][]{Mason01}\footnote{Values
  are listed from the 03 Mar 2013 update of WDS.} as WDS 22481-2422
and with discovery identifier ``WSI 138''\footnote{WSI = Washington
  Speckle Interferometer.}. A single observation is reported for epoch
2010, with a reported companion at separation 0''.5 at PA =
144$^{\circ}$, with magnitudes 12.80 and 14.80 (presumably $V$-band,
as the combined magnitude [12.64] is similar to the adopted $V$
magnitude in Table \ref{tab:phot}).  We are unable to confirm the
existence of the companion reported in the WDS. In the 118 frames
taken during 25 nights, with FWHMs in the range 1''.2 to 2''.8, LP
876-10 appeared to be a point source - with no evidence of
elongation. With only a single observation, the possibility remains
that the reported WDS companion may be a chance alignment between this
high proper motion star and a background star (B. D. Mason 2013,
private communication). However, we believe that a background star is
unlikely to explain this discrepancy. Based on the UCAC4 position of
LP 876-10 for epoch 2000.0 \citep{Zacharias13}, the proper motion
calculated in this paper, and the separation/PA value listed in WDS,
we estimate that the WSI 138 companion reported in WDS had approximate
ICRS position 22:48:04.76 -24:22:09.1 (epoch 2010). The only object
listed in any Vizier-queryable catalog within 2'' of this position is
the WISE detection of LP 876-10 itself (0.5'' away) during 2010. No
plausible optical-IR counterpart within 2'' of this position exists in
the USNO-B1.0, SuperCOSMOS, GSC, and 2MASS catalogs. It seems very
unlikely that a bright (V = 14.8) background star can explain the
faint companion to LP 876-10 reported in the WDS. If the companion
were real, and physically associated with LP 876-10, then its absolute
magnitude (M$_V$ = 15.40) would correspond to a 0.11 \msun\, star on
the calibration of \citet{Delfosse00}. Given the projected separation
(0''.5 = 3.8 AU), these values would predict an orbital period of
$\sim$13.5 yr. Assuming zero eccentricity and face-on projection, one
would predict orbital motion of $\sim$27 deg\,yr$^{-1}$ and a
photocentric amplitude of $\sim$110 mas.\\

\begin{figure}[htb!]
\includegraphics[scale=0.35]{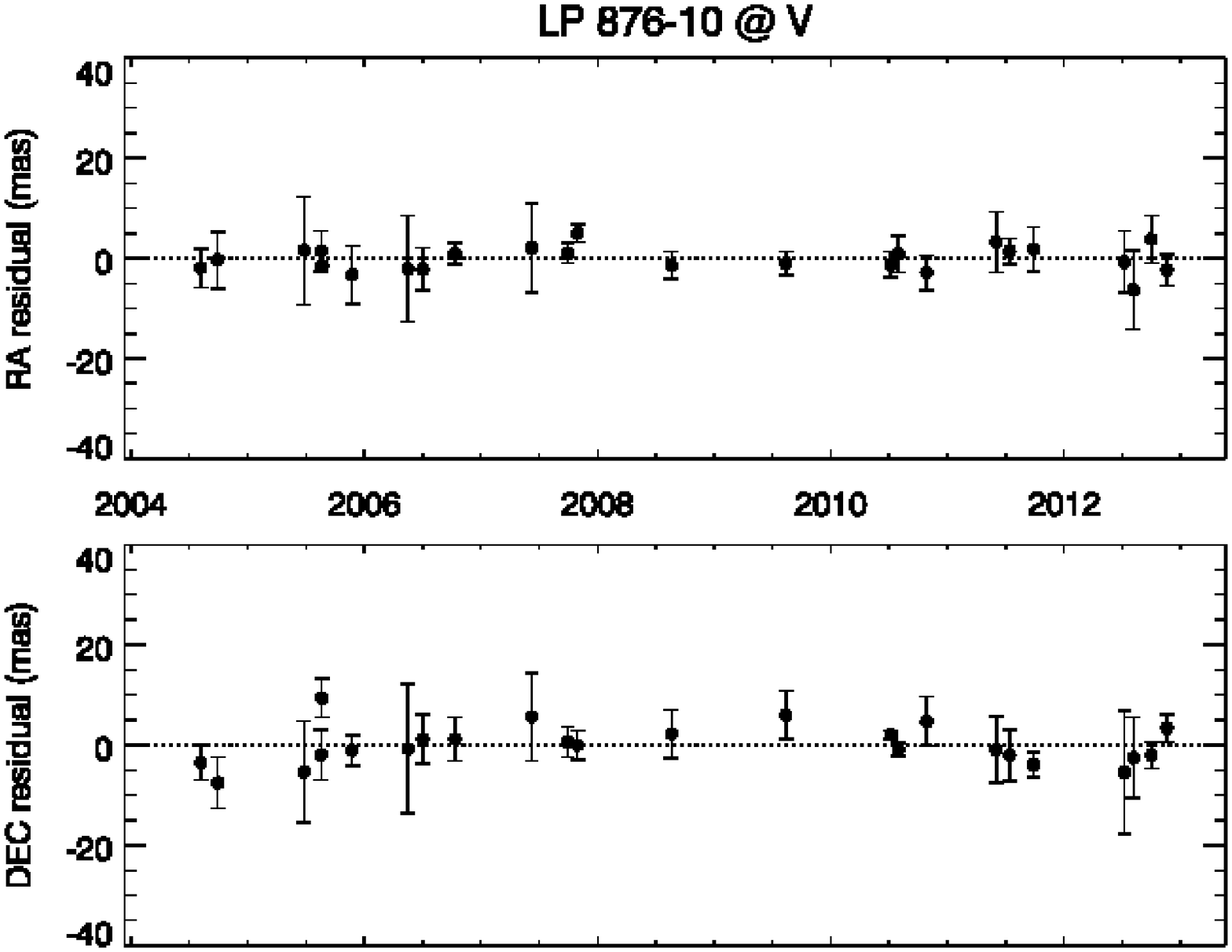}
\caption{Astrometric residuals in RA ({\it top}) and Dec ({\it
    bottom}) for LP 876-10 in V-band images after subtracting the
  parallactic motion with $\varpi$ = 132.07\,$\pm$\,1.19 mas and
  proper motion $\mu$ = 378$^{\circ}$.1\,$\pm$\,0.4 \masyr\, at PA
  118$^{\circ}$.0\,$\pm$\,0.1 east of north. If the companion reported
  in the WDS (WSI 138 B) were real, we predict that it would produce
  an astrometric perturbation on the photocenter at the $\sim$110 mas
  level with period $\sim$13 yr. Any perturbations due to unseen
  companions must be at the $<$10 mas level over the $\sim$8 yr
  baseline.
\label{fig:astrom}}
\end{figure}

The predicted photocentric amplitude would be about half (50 mas over
8 years) of the full amplitude (110 mas over $\sim$13.5 years) during
the observations to date. As seen in Fig. \ref{fig:astrom}, the
astrometric solution using only parallax and proper motion is quite
good, and any gravitational perturbations on LP 876-10 must be at the
$<$10 mas level over $\sim$8 yr, which easily rules out the predicted
signal for the companion reported in the WDS.  Table \ref{tab:pm}
shows that the difference between the long-term proper motions
(e.g. SuperCOSMOS, USNO-B1.0, PPMX, UCAC4) are largely within
$\sim$5-10 \masyr\, (rms) of the 8-year baseline proper motion
calculated in this survey, further suggesting that it would be
difficult to hide a $\sim$50 \masyr\, perturbation of the photocentric
motion. As the purported WDS companion should have a period only
somewhat longer than the duration of our RECONS astrometric dataset,
and with a predicted photocentric amplitude similar in size to the
observed parallax, we conclude that it is unlikely that the companion
reported in the WDS catalog is real.\\


\subsection{Temperature, Luminosity, and Radius\label{sec:Teff}}

We estimated \teff\, for LP 876-10 by fitting the photometry in Table
\ref{tab:phot} to the BT-Settl grid of synthetic stellar spectra which
vary by effective temperature, metallicity, and surface gravity
\citep{Allard12}. Twenty-two colors consisting of combinations of the
bands $V$, $R_{RC}$, $I_{RC}$, $J$, $H$, $K_s$, $W1$, $W2$, and $W3$
were compared to grid interpolations based on models, and the best fit
yielded an interpolated temperature of \teff\, = 3132\,K and solar
metallicity. We estimated the uncertainty in \teff\, due to
metallicity and surface gravity by individually varying these
parameters by one increment (0.2 dex) and measuring the effect on the
resultant \teff. The uncertainty in the \teff\, breaks down
approximately as follows: $\pm$33\,K from the dispersion in
color-based \teff\, estimates for the best fit, $\pm$50\,K due to
metallicity uncertainty, and $\pm$25\,K due to uncertainty in
log($g$). Together this yields an overall \teff\, uncertainty of
$\pm$65~K. The systematic error due to the validity of the BT-Settl
models is unknown, however our derived \teff\, should be comparable to
M dwarf \teff\, values derived using the same models \cite[indeed
][similarly derives \teff\, $\simeq$ 3100-3200\,K for M4 dwarfs like
  LP 876-10 using BT-Settl models]{Rajpurohit13}. The best fitting
BT-Settl synthetic spectrum had \teff\,= 3100\,K, [Fe/H]\,=\, 0.0,
log($g$)\,=\, 5.0. From considerations of the star's color-magnitude
diagram position (Sec. 2.6), we predict that LP 876-10 has a slightly
subsolar metallicity, and lies near the zero-age main sequence for
$\sim$0.2 \msun\, stars \citep[log($g$) $\simeq$ 5.06; ][]{Baraffe98}.
The best fitting BT-Settl synthetic spectrum then adjusted via an
iterative process to produce a match to the observed photometry.  The
process determines a small $\lambda$-dependent polynomial correction
factor that is applied to the synthetic spectrum to cause small
modifications in order to produce the best fit to the photometry
\citep[details of the technique are described in][]{Dieterich13}.\\

By directly integrating the spectral energy distribution made by
fitting the photometry in Table \ref{tab:phot} with solar composition
BT Settl models, we estimate \mbol\, = 9.994\,$\pm$\,0.020, luminosity
= (1.763\,$\pm$\,0.042) $\times$ 10$^{31}$ \ergs, \logl\, =
-2.337\,$\pm$\,0.010, absolute bolometric magnitude \Mbol\, =
10.597\,$\pm$\,0.026, and bolometric correction BC$_V$ = \mbol\, - V =
-2.62\,$\pm$\,0.02 \citep[adopting solar parameters from
][]{Mamajek12, Pecaut13}.  Combining this luminosity with our previous
\teff\, estimate, we estimate a radius of 0.23\,$\pm$\,0.01 \rsun.
Combined with our estimate of the projected rotation velocity \vsini\,
(22\,$\pm$\,2 \kms), this places an upper limit on the rotation period
of LP 876-10 of 0.55\,$\pm$\,0.05 day (see \S2.8).\\

\subsection{Color-Magnitude Diagram and Metallicity \label{sec:absmag}} 

Using our new parallax and the photometry in Table \ref{tab:phot}, we
estimate absolute magnitudes of \mv\, = 13.21\,$\pm$\,0.02 and M$_{\rm
  K_s}$ = 7.81\,$\pm$\,0.03.  From Table 2, we calculate a \vk\, color
of 5.40\,$\pm$\,0.02 mag.  Using the \vk\, vs. \mv\, relations from
\citet{Henry04} and \citet{Johnson09}, we predict photometric
distances of 7.9\,$\pm$\,1.5 pc and 7.7\,$\pm$\,1.4 pc, respectively,
in excellent agreement with our trigonometric parallax distance.  The
agreement between the trigonometric parallax distance and the
available photometric distances \citep[][this section]{Reid03} is also
indicative that LP 876-10 is unlikely to have an unresolved companion
of similar mass, and it is more likely to be a main sequence, rather
than pre-main sequence, star.\\

We can constrain the metallicity and age using the star's
color-magnitude data.  In Fig. \ref{fig:vk}, we plot the star's \vk\,
color versus absolute magnitude \mv\, and use the metallicity
color-magnitude calibration of \citet{Johnson09} to predict a
metallicity of [Fe/H] = -0.07 dex (estimated accuracy $\pm$0.06 dex).
The calibration of \citet{Schlaufman10} predicts a metallicity of
[Fe/H] = -0.15 dex.  These are in reasonable agreement with the high
S/N estimate for TW PsA (Fomalhaut B) from \citet{Barrado97} ([Fe/H] =
-0.11\,$\pm$\,0.02). Other published [Fe/H] estimates for TW PsA are
-0.01\,$\pm$\,0.09 \citep{Santos04} and -0.20 \citep{Morell94}.
Hence, both the photometric metallicity estimate for LP 876-10 and the
spectroscopic metallicity estimates for TW PsA are self-consistent,
and consistent with being very slightly subsolar [Fe/H] $\simeq$ -0.1
dex.\\

\subsection{Mass and Age Constraints}

\begin{figure}[htb!]
\includegraphics[angle=0,scale=0.5]{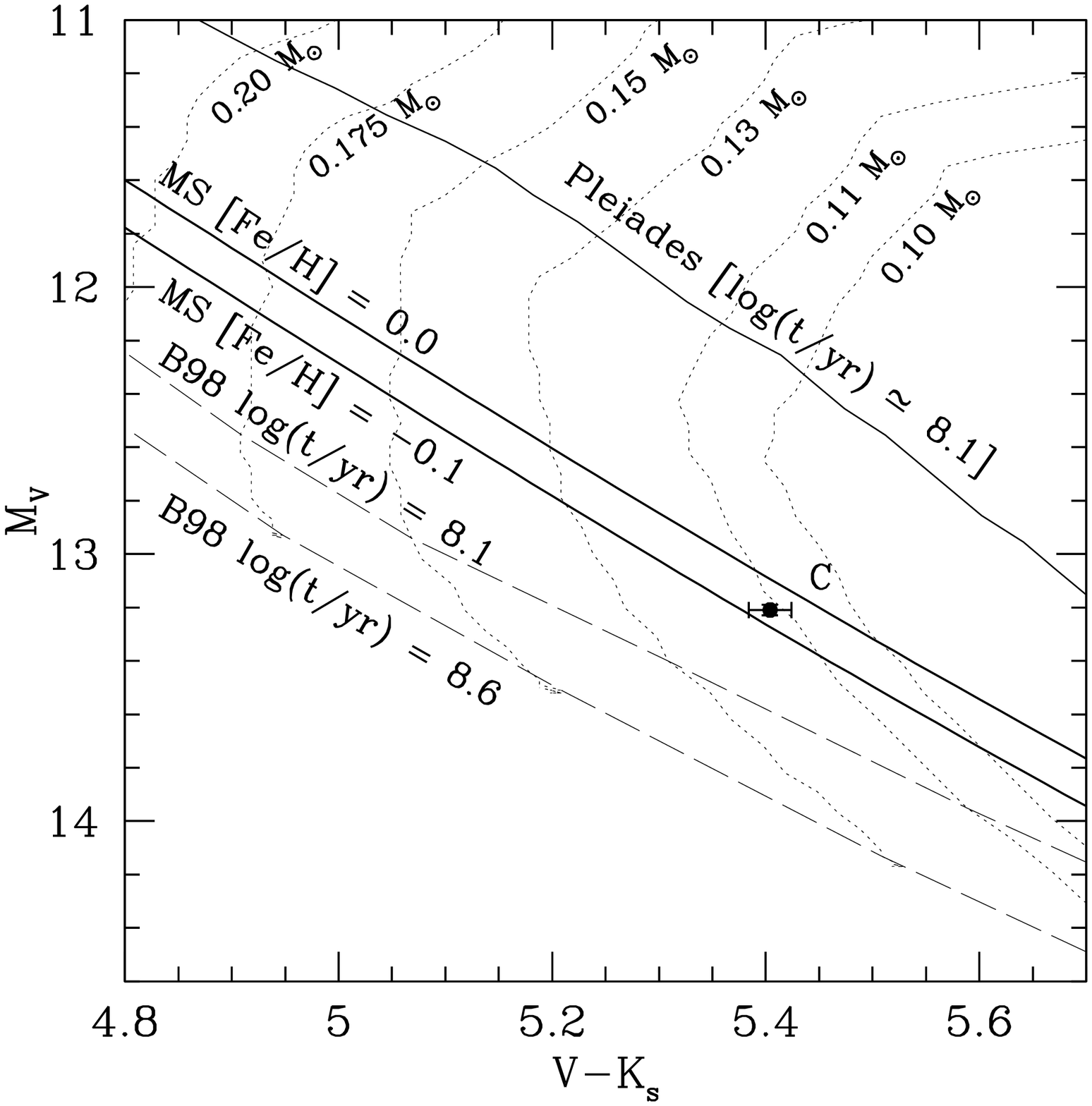}
\caption{Color-magnitude (V-K vs. \mv) for Fomalhaut C (LP 876-10).
  The {\it empirical} main sequences for [Fe/H] = 0.0 and -0.1 from
  \citet{Johnson09} are plotted as {\it thick solid lines}, along with
  the color-magnitude sequence for the $\sim$125 Myr-old Pleiades
  cluster from \citet[][adopting $d$ = 133.5 pc from Soderblom et
  al. 2005]{Stauffer07} ({\it thin solid line}). The {\it theoretical}
  evolutionary tracks from \citet{Baraffe98} are plotted as {\it
    dotted lines}, and the isochrones for the approximate age of the
  Pleiades (log(t/yr) = 8.1) and Fomalhaut A \& B (log(t/yr) = 8.6)
  are plotted as {\it long dashed lines}. The evolutionary tracks do
  not accurately predict the solar composition main sequence nor
  Pleiades sequence for this color-magnitude combination. Note that
  the Pleiades has a well-determined Lithium depletion boundary and
  main sequence turn-off age consistent with $\sim$125 Myr
  \citep{Ventura98, Stauffer98, Barrado04}, and the main sequence
  ``turn-on'' appears to be consistent with this age as well
  \citep{Barenfeld13}. Fomalhaut C appears to lie on the empirical
  main sequence of \citet{Johnson09} with [Fe/H] $\simeq$ -0.07.
\label{fig:vk}}
\end{figure}

Using the \citet{Delfosse00} \mv\, vs. mass calibration for field M
dwarfs (i.e. mixed metallicities and ages), the approximate mass of LP
876-10 is $\sim$0.20\,\msun. Interpolating within the
\citet{Baraffe98} tracks, one finds that solar composition stars with
masses of greater than 0.163 M$_{\odot}$ are not ever predicted to be
as faint as M$_V$ = 13.21 mag (see Fig. \ref{fig:vk}). As the tracks
are first and foremost tracing luminosity evolution as a function of
mass and age, we also examine the constraints that the luminosity of
LP 876-10 can provide. Through fitting BT-Settl models to the
photometry, we estimate the luminosity to be \logl\, =
-2.337\,$\pm$\,0.010 dex.  We find that the \citet{Baraffe98} {\it
  and} \citet{Dotter08} solar composition tracks give essentially
identical predictions that no stars with with masses greater than
0.197 M$_{\odot}$ are predicted to have luminosities this low.  Using
those tracks, we estimate that it takes a 0.2\,\msun\, star
approximately $\sim$300 Myr to reach within $\sim$0.01 mag of the
zero-age main sequence (the actual minimum in luminosity and radius
occurs around $\sim$400-500 Myr). The appearance of LP 876-10 on the
zero-age main sequence for [M/H] $\simeq$ -0.1 is commensurate with
the adopted age for Fomalhaut A and B \citep[440
Myr;][]{Mamajek12}. 

As seen in Fig. \ref{fig:vk}, the \citet{Baraffe98} isochrones do not
accurately reproduce the empirical main sequence from
\citet{Johnson09} in this color regime, so our lower bound on the age
of LP 876-10 is only approximate.  Naively interpolating the mass and
age of Fomalhaut C from the evolutionary tracks and isochrones would
yield a mass of $\sim$0.11 \msun\, and age of $\sim$60 Myr. However,
as can be seen in Fig. \ref{fig:vk}, a 125 Myr isochrone (log(age/yr)
= 8.1) from the same tracks fails to replicate the intrinsic
color-magnitude sequence for the $\sim$125 Myr-old Pleiades
\citep{Barrado04}. For the V-K$s$ color (5.4) of LP 876-10, the
combination of Pleiades color-magnitude sequence from
\citet{Stauffer07} and mean Pleiades distance from \citet{Soderblom05}
yield a Pleiades absolute magnitude of \mv\, = 12.24. The
\citet{Baraffe98} isochrones for age 125 Myr (log(age/yr) = 8.1)
predict absolute magnitude \mv\, = 13.57 for V-K$_s$ =
5.4\footnote{The \citet{Baraffe98} tracks use the CIT $JHK$
  photometric system. We convert the \citet{Baraffe98} CIT photometry
  to 2MASS following \citep{Carpenter01}.}, i.e.  1.33 mag too faint!
As summarized by \citet{Bell12}, ``for all optical colours, no pre-MS
models follows the observed Pleiades sequence for temperatures cooler
than 4000\,K.'' Estimating isochronal ages using pre-MS evolutionary
tracks is quite problematic, with large systematic differences between
tracks \citep[see review by][]{Soderblom10}. For all of these reasons,
we do not adopt the pre-MS mass and isochronal age interpolated from
the evolutionary tracks and isochrones in Fig. \ref{fig:vk}, and
instead constrain the age based on its proximity to the main sequence,
and infer the mass based on main sequence absolute magnitude vs. mass
considerations.  Given the empirical and theoretical constraints
previously discussed, we adopt a mass of 0.18\,$\pm$\,0.02 \msun\, for
Fomalhaut C.\\


\subsection{Rotation Period}

\begin{figure}[htb!]
\includegraphics[angle=-90,scale=0.6]{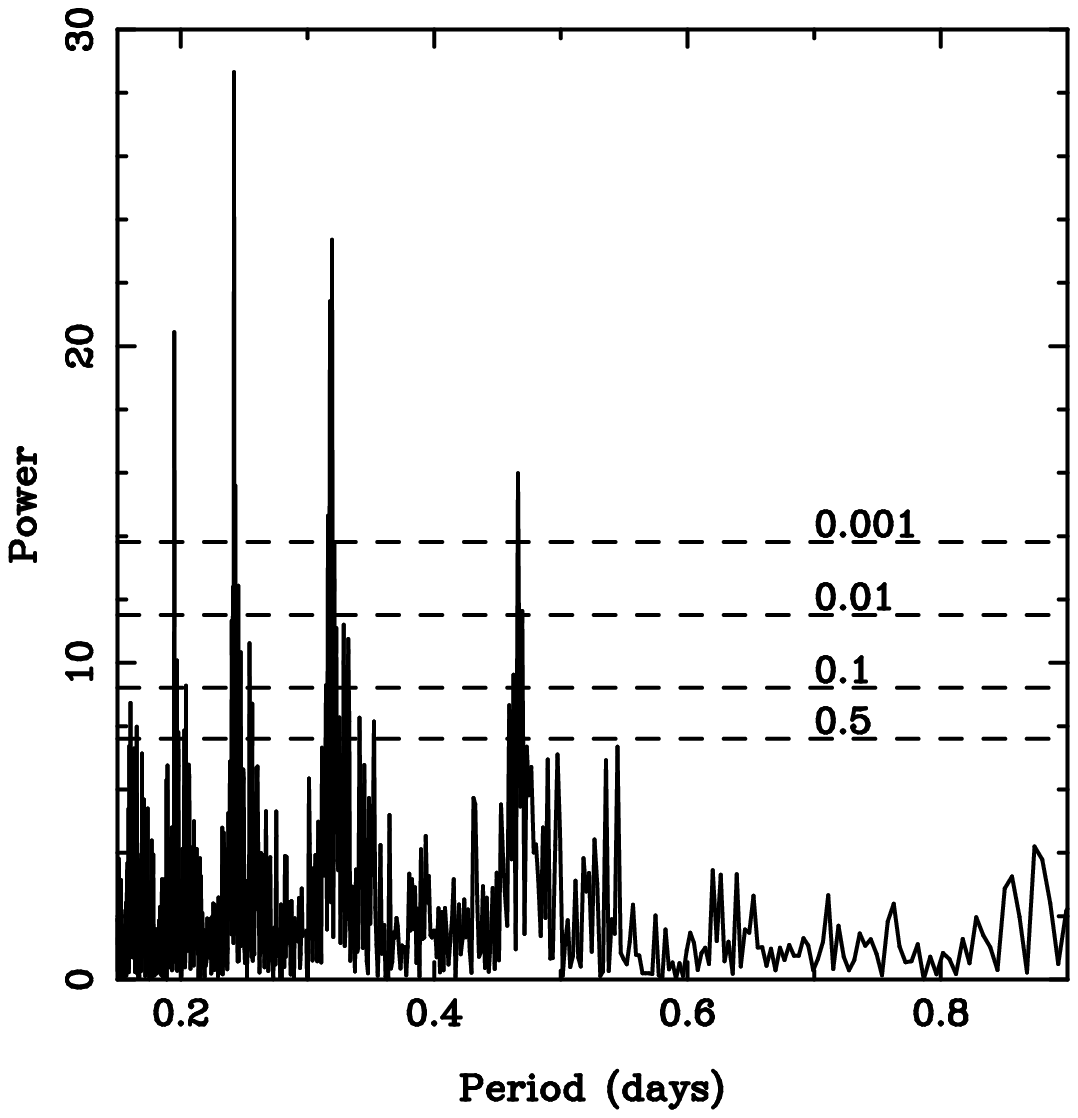}
\caption{Lomb-Scargle periodogram for SuperWASP photometry for LP
  876-10.  The period at 0.466 d appears to be the real period, and
  tests indicate that the periods at 0.242 and 0.318 d are due to
  aliasing.
\label{fig:periodogram}}
\end{figure}

Photometric data from the online SuperWASP
archive\footnote{http://www.wasp.le.ac.uk/public/} \citep{Butters10}
consisting of 14,991 measurements for LP 876-10 were extracted for two
observing seasons (2007-2008). To search for a rotation period, we
selected SuperWASP photometry from a single well-sampled season (2008)
with $V_{SuperWASP}$ magnitudes between 12.46 and 12.70, with
magnitude and photometric error of less than 0.2 mag, and with a good
TAMFLUX2 flag extraction.  SuperWASP photometry is calibrated to the
Tycho-2 $V_T$ system \citep{Pollacco06,Hog00}.  There were 3162 points
for subsequent analysis.  To remove 1-day aliasing effects, all points
during a single observing night were adjusted so that their average
equalled the average seasonal magnitude of LP 876-10. A Lomb-Scargle
periodogram with associated False Alarm Probabilities (FAP) was
calculated following \citet{Press92}, and the resultant periodogram is
plotted in Fig. \ref{fig:periodogram}. There is significant power (FAP
$<$ 0.001) seen in the LS periodogram of LP 876-10 at periods of
0.195, 0.242, 0.318 and 0.466 days. A period of P\,=\,0.466 d would
correspond to an equatorial velocity of 26 \kms, which is only
slightly larger than the observed \vsini\, (22\,\kms, corresponding to
a maximum period of 0.55 day). For the star's mass and radius, we
estimate a breakup velocity and period \citep[following][]{Townsend04}
of 386 \kms\, and P$_{breakup}$ = 0.03 day, respectively. Hence, any
of the periods between $\sim$0.03 and $\sim$0.55 day are possible,
given the breakup and \vsini\, constraints, respectively.  The fastest
rotation period among 41 nearby field M dwarfs in the MEarth survey of
\citet{Irwin11} is 0.28 days. We test the robustness of the detection
by injecting artificial sinusoidal ($P = 0.466$ day) signals into a
Gaussian distributed photometric data set with the same time cadence
as the LP 876-10 data set. These tests indicate that the 0.195, 0.242,
and 0.318 day peaks are aliasing effects due to the irregular time
sampling of the light curve. We conclude that the $P = 0.466$ d peak
is most likely due to the rotation of the star.\\

Unfortunately, a rotation period of $\sim$0.5 day for a $\sim$0.2
\msun\, star places negligible constraint on its age. Mid-M stars with
rotation periods faster than 1 day are a nearly ubiquitous feature of
stellar samples between ages of $\sim$2 Myr and $\sim$10 Gyr
\citep[see Fig. 12 of ][]{Irwin11}.  Figure 11 of \citet{Irwin11}
plots the masses of field M dwarfs vs. their rotation periods measured
by the MEarth survey. For stars of $\sim$0.2 \msun, a rotation period
of $\sim$0.466 day is fast, however not unprecedented. Indeed,
\citet{Irwin11} finds that mid-M dwarfs like LP 876-10 can have
periods of less than 1 day whether they are thin disk or thick disk
stars. The survey of \citet{Irwin11} had little difficulty finding
kinematically old ($>$7 Gyr) thick disk M dwarfs with sub-day rotation
periods. We conclude that attempts to age-date LP 876-10 via
gyrochronology/rotation constraints appear fruitless.


\subsection{Activity}

Not only is LP 876-10 fast rotating, but, unsurprisingly, it appears
to be a coronally active star as well. \citet{Voges99} ranked LP
876-10\footnote{Listed under its Guide Star Catalog alias
  ``GSC6964.01226''.} as the most likely optical counterpart of the
ROSAT All-Sky Survey (RASS) Bright Source Catalog (BSC) X-ray source
1RXS J224803.5-242240. The X-ray counterpart is 35'' away from LP
876-10.  However, the RASS BSC position error is 15'', and LP 876-10
is the brightest optical source within 40'', indicating that it is the
likely X-ray source \citep{Neuhauser95}. 1RXS J224803.5-242240 appears
to be the brightest RASS X-ray source within a degree of LP
876-10. The fact that the position of the brightest RASS X-ray source
within a degree of LP 876-10 lies within 40'' of the rapidly rotating,
nearby M dwarf suggests to us that LP 876-10 is almost certainly the
optical counterpart of 1RXS J224803.5-242240. The RASS-BSC catalog
\citep{Voges99} reports a soft X-ray flux of 0.142 ct\,s$^{-1}$ (28\%
uncertainty) with HR1 hardness ratio of -0.23$\pm$0.21, detected over
a short exposure time of 176 s. Using the energy conversion factor
relation from \citet{Fleming95}, this translates to a coronal X-ray
flux in the soft X-ray band (0.2-2.4 keV) of roughly
1.01$\times$10$^{-12}$ erg/s/cm$^{-2}$. At $d$ = 7.57 pc, this
corresponds to an X-ray luminosity of L$_X$ $\simeq$ 10$^{27.84}$
erg/s. This implies \loglxlbol\, $\simeq$ -3.41, i.e. a very active
star close to X-ray saturation. This corroborates the very high
projected rotational velocity measured spectroscopically (\vsini\, =
22 \kms), which should induce strong magnetic activity.\\

\subsection{Velocity and Interloper Probability}

With our best measurements of the proper motion, radial velocity, and
parallax, we calculate the 3D Galactic velocity of LP 876-10 to be (U,
V, W) = -5.3\,$\pm$\,0.2, -7.6\,$\pm$\,0.3, -11.9\,$\pm$\,0.4
\kms. Comparing these values to those for Fomalhaut and Fomalhaut B (TW
PsA), we find that LP 876-10's velocity only differs from that of
Fomalhaut by 1.1\,$\pm$\,0.7\,\kms, and that of Fomalhaut B by
1.1\,$\pm$\,0.5\,\kms.  Using the LSR velocity ellipsoid for both dM
and dMe dwarfs estimated by \citet[][their unweighted
  solution]{Reid02}, and adopting the solar peculiar velocity with
respect to the LSR from \citet{Schonrich10}, we naively only expect
roughly 1 in $\sim$55,000 field M dwarfs to have UVW velocities within
1.1 \kms\, of Fomalhaut, and roughly 1 in $\sim$12,000 field M dwarfs
to have a velocity within 2 \kms.\\

\citet{Henry06} report 239 M dwarfs within 10 pc having accurate
trigonometric parallaxes. These numbers are updated at recons.org,
with a count as of 1 January 2012 of 248, which corresponds to a
number density of 0.059 pc$^{-3}$. This space density implies that
within a sphere of radius 1 pc surrounding Fomalhaut, we would expect
to find 0.25 M dwarfs. Hence, we estimate the probability that a
random M dwarf could appear within 1 pc of Fomalhaut, and sharing its
velocity within less than 2 \kms, as approximately 1 in
$\sim$10$^{4.7}$ (and sharing its velocity within less than 1.1 \kms\,
as roughly 1 in $\sim$10$^{5.3}$).  For comparison, one would expect
to have to encircle a sphere $\sim$36 pc in radius in the local
Galactic disk in order to find another M dwarf whose motion randomly
agreed with that of Fomalhaut within less than 2\,\kms.  Our
probability estimates do not take into account the similarity in the
spectroscopic metallicity of TW PsA and the photometric metallicity of
LP 876-10, which provides further agreement. We conclude that LP
876-10 appears to be related to Fomalhaut and TW PsA beyond a
reasonable doubt.\\

\subsection{The Castor Moving Group}

Fomalhaut was listed by \citet{Barrado98} as a potential member of the
Castor Moving Group (GMG). The co-motion of LP 876-10 with Fomalhaut
may be less significant if Fomalhaut is immersed in a swarm of
co-moving stars like the purported CMG. The origin and nature of
moving groups like CMG is an active field of study
\citep[e.g.][]{Famaey05, Murgas13}. That the CMG represents a
kinematic group of stars of common age and birthsite is unlikely.

Calculating revised space motions for the 14 CMG ``members'' (``Y'' or
``Y?'' members) from \citet{Barrado98}, using revised Hipparcos
astrometry \citep{vanLeeuwen07} and the best available radial
velocities \citep{BarbierBrossat00, Gontcharov06}, we find that the
CMG stars have median velocity of ($U$, $V$, $W$) = -11.1\,$\pm$\,1.9,
-8.6\,$\pm$\,0.8, -9.7\,$\pm$\,1.0 \kms, with standard deviations of
6.1, 3.6, and 4.2 \kms. The scatters are much larger than the typical
velocity errors, and larger than the one-dimensional velocity
dispersions of nearby clusters and associations \citep[$<$1.5
  \kms;][]{Madsen02, Mamajek10}. The velocity for Fomalhaut differs
from the CMG median velocity by 5.6\,$\pm$\,2.3 \kms.  The list of
``final'' members in \citet{Barrado98} comprises $\sim$27 \msun\, of
stars spread out over a volume of $\sim$55,000 pc$^3$, implying that
the density of CMG members in the solar neighborhood is roughly
$\sim$0.004$\times$ the local disk density \citep[0.12
  \msun\,pc$^{-3}$; ][]{vanLeeuwen07}. The stellar systems in the CMG
have negligible interaction with one another, and so their motions are
completely dominated by the local Galactic potential.\\

The velocity differences between Fomalhaut and individual CMG members
is illuminating, and we discuss the famous CMG members Vega, LP
944-20, and Castor itself, in more detail. Vega is a proposed fellow
CMG star of either similar age \citep[455\,$\pm$\,13 Myr;][]{Yoon10}
or somewhat older age \citep[700$^{+150}_{-75}$ Myr;][]{Monnier12}
than that of Fomalhaut \citep[440\,$\pm$\,40 Myr;][]{Mamajek12}. Could
Vega and Fomalhaut be related?  Using the revised Hipparcos astrometry
for Vega and its mean radial velocity reported by
\citet{Parthasarathy87}, we estimate for Vega a velocity of ($U$, $V$,
$W$) = -15.9\,$\pm$\,0.7, -6.2\,$\pm$\,0.5, -7.7\,$\pm$\,0.3
\kms. Vega's velocity differs from that of Fomalhaut by
10.9\,$\pm$\,1.0 \kms, and only 10 Myr ago their separations differed
by $\sim$110\,$\pm$10 pc. Another nearby famous CMG ``member'' is the
nearby candidate brown dwarf LP 944-20\footnote{A new RECONS parallax
  has been measured which places LP 944-20 at a distance of 6.4 pc
  (Dieterich et al., submitted), making it most likely a star near the
  H-burning limit rather than a brown dwarf. The new distance revises
  LP 944-20's space motion to ($U$, $V$, $W$) = -14.9, -5.9, -1.5
  \kms, which differs from that of Fomalhaut by 13.5 \kms, and does
  not change the qualitative conclusions in the text.}
\citep{Ribas03}. Adopting the astrometry from \citet{Tinney96} and a
mean radial velocity of +9.0\,$\pm$\,0.5 \kms\, \citep[based on
  measurements from ][]{Martin06}, we calculate a velocity for LP
944-20 of ($U$, $V$, $W$) = -12.2\,$\pm$\,0.4, -5.6\,$\pm$\,0.3,
-2.8\,$\pm$\,0.3 \kms.  LP 944-20 is currently situated 6.6 pc away
from Fomalhaut, and its velocity differs from that of Fomalhaut by
10.9\,$\pm$\,0.8 \kms. Only 10 Myr ago, LP 944-20 and Fomalhaut were
separated by $\sim$100\,$\pm$\,8 pc, and were only more widely
separated in the past. We also investigated whether there was any
association between Fomalhaut and Castor itself.  For the Castor
sextuplet system, we adopt the recent parallax estimate from
\citet{Torres02}, the long-term system proper motion from PPMX
\citep{Roeser08}, and the center-of-mass radial velocity estimate from
\citet[][we adopt a RV uncertainty of 1 \kms\, in our
  calculations]{Heintz88}. These values are consistent with the Castor
system having a velocity of ($U$, $V$, $W$) = -7.5\,$\pm$\,0.7,
-3.7\,$\pm$\,0.6, -11.5\,$\pm$\,0.4 \kms. Fomalhaut is currently
$\sim$21 pc away from the Castor system, differing in velocity by a
significant margin (4.9\,$\pm$\,1.1 \kms), and only 10 Myr ago
Fomalhaut and Castor were separated by $\sim$50\,$\pm$\,5 pc, and were
even more distant in the past (more than 700 pc 100 Myr ago).\\

Despite these stars (the Fomalhaut system, Vega, LP 944-20, Castor
system) being young and having somewhat similar velocities, their
velocities are well-constrained enough and different enough that it is
clear that they were not in the vicinity of one another even in the
recent past, let alone a couple of Galactic orbits ago.  We conclude
that the Castor Moving Group is comprised of stars from different
birthsites rather than a coeval system, and hence ``membership'' to
the CMG does not provide useful age constraints for the Fomalhaut
system (or Vega, LP 944-20, Castor, or other CMG members).\\

\subsection{A Bound Companion?}

One predicts that stellar companions in multiple systems can exist
with separations up to their tidal (Jacobi) radius with respect to the
Galactic potential.  \citet{Jiang10} parameterize the tidal radius
$r_{t}$ as:

\begin{equation}
r_{\rm t}\,= \left\{ \frac{G(M_1 + M_2)}{4\Omega A} \right\}^{1/3}
\end{equation}

\noindent where G is the Newtonian gravitational constant, M$_1$ and
M$_2$ are the masses of the stars, $\Omega$ is the Galactic angular
circular speed (orbital velocity dividied by Galactocentric radius),
and $A$ is the Oort parameter.  Adopting modern estimates of the
relevant Galactic parameters, and rewriting the expression from
\citet{Jiang10}, we estimate the tidal radius to be:

\begin{equation}
r_{\rm t}\,=\,1.35\,{\rm pc}\,\left\{ \frac{M_{\rm total}}{M_{\odot}} \right\}^{1/3}
\end{equation}

\noindent Summing the masses of the Fomalhaut system components (2.83
\msun), one predicts a tidal radius of $\sim$1.9 pc. The tidal radius
for Fomalhaut A alone is $\sim$1.7 pc. Hence, the separation of
$\sim$0.8 pc between Fomalhaut C and the A is not dynamically
implausible for a bound system. Recent systematic surveys for
wide-separation pairs using modern astrometric databases have started
to yield many previously unrecognized parsec-scale common proper
motion pairs \citep[e.g.][]{Shaya11}, making Fomalhaut C less unusual
than it may have appeared even a decade ago. Stable orbits for time
scales longer than a Gyr are also possible for separations larger than
the tidal radius if the distant companion orbits retrograde to the
Galactic rotation \citep[e.g.][]{Makarov12}. Indeed, higher
precision-radial velocities for the components of the Fomalhaut system
and taking into account the sub-kilometer second$^{-1}$ effects of
convective blueshift and gravitational redshift, may lend itself to
providing a test as to whether Fomalhaut C is orbiting Fomalhaut AB
either retrograde or prograde to the Galactic rotation (V. Makarov,
priv. comm.).\\

Fomalhaut A and B are separated by $\Delta_{AB}$ =
57.4$^{+3.9}_{-2.5}$ kAU, and Fomalhaut C is separated by
$\Delta_{AC}$ = 158.2$^{+2.3}_{-1.2}$ kAU from A, and by $\Delta_{BC}$
= 203.4$^{+1.0}_{-0.8}$ kAU from B. We calculate the position of the
barycenter (center of mass) for the system using the Galactic (X, Y,
Z) positions and masses in Table 1: ($X, Y, Z$)$_{\rm com}$ = 3.08,
1.13, -6.93 pc. Converting this position to the equatorial ICRS
coordinate system yields ($\alpha$, $\delta$) = 344$^{\circ}$.179,
-29$^{\circ}$.792, at distance 7.67 pc.  We can make a rough estimate
of the orbital period of C around the AB pair. C is currently located
$\sim$0.77 pc from the system's center of mass. If C is currently near
apastron (not an unreasonable assumption given that binary stars will
spend most of their time near apastron), and if C's periastron must
almost certainly be larger than B's current separation from the system
barycenter (0.24 pc), then a reasonable first estimate of C's orbit is
$a$ $\sim$ 0.5 pc and $e$ $\sim$ 0.5. For the total mass of the
Fomalhaut system (2.83 \msun), this translates to an approximate
orbital period of $\sim$20 Myr, or $\sim$5\%\, the system's
  age. The predicted orbital velocity of LP 876-10 around the
  Fomalhaut system barycenter would be $\sim$0.15 \kms. Given the
  masses and configuration of the AB pair, the escape velocity of C is
  $\sim$0.2 \kms.\\

How stable is Fomalhaut C's orbit with respect to A and B? Obviously,
the orbit of AB and AB-C are not well constrained. We only have fairly
accurate estimates of the relevant mass ratios and current
separations, while the semi-major axes and eccentricities are
unknown. The mass of C is very small compared to that for the AB pair
($\mu$ = $M_C$/($M_A$ + $M_B$) $\simeq$ 0.07), and its current
separation from the center of mass for the system is approximately 159
kAU. Based on simulations of test particles in the vicinity of binary
systems of varying semi-major axis, mass ratio, and eccentricity,
\citet{Holman99} provided estimates of the widest stable orbit around
a member of a binary system (S-type orbits), and the closest orbit
around both members of a binary system (P-type orbits). If the current
A-B separation is equivalent to its semi-major axis (assume $e$ = 0;
$a$ = 57.4 kAU), then the minimum stable semi-major axis for C is
predicted to be $\sim$135 kAU. \citet{Tokovinin98} estimates that the
mean eccentricity for wide binary pairs is $<e>$ $\simeq$ 2/3. If
Fomalhaut B is currently near apastron with $e$ $\sim$ 2/3, then $a$
$\sim$ 34 kAU, and the minimum stable semi-major axis for Fomalhaut C
is $\sim$140 kAU. There are plausible ranges of orbital parameters for
Fomalhaut B and C that would be dynamically stable over many orbits.\\

Could LP 876-10 be genetically related to Fomalhaut AB but we are
``catching it in the act'' of being an unbound escapee of the
Fomalhaut system? We argue that this is very unlikely. LP 876-10 has
velocity statistically consistent with that of Fomalhaut A and B
($\Delta$$S$ = 1.1\,$\pm$\,0.7 \kms). If the star actually had a
velocity difference of $>$0.2 \kms\, (i.e. above escape velocity),
with respect to the Fomalhaut AB barycenter, then it would not spend
much time in the vicinity of Fomalhaut or near its tidal radius. The
approximate timescale that LP 876-10 would spend within Fomalhaut's
tidal radius is approximately $t$ $\simeq$ $r_t$/$\Delta S$, where
$r_t$ $\simeq$ 1.9 pc is the tidal radius of Fomalhaut AB, and we
posit that $v$ must be larger than the escape velocity (0.2 \kms\, =
0.2 pc Myr$^{-1}$). Hence:

\begin{equation}
t_{Myr}\, \simeq\, \frac{r_t}{\Delta S} < 9.5 {\rm Myr}
\end{equation}

For $\Delta S$ $\sim$ $v_{esc}$ $\sim$ 0.2 \kms, LP 876-10 could spend
of order $\sim$10 Myr within the tidal radius of Fomalhaut. For a
velocity difference of $\sim$1 \kms, LP 876-10 would spend only
$\sim$2 Myr. Velocity differences between LP 876-10 and Fomalhaut of
$\Delta S$ greater than 2.5 \kms\, are ruled out at 95\%\, confidence,
so timescales for LP 876-10 being unbound and within the tidal radius
of Fomalhaut shorter than $\sim$0.8 Myr are ruled out. Hence, $t$
would have to be of order $\sim$1-10 Myr if LP 876-10 is unbound to
Fomalhaut AB.  For a main sequence lifetime of Fomalhaut A of
$\sim$0.9 Gyr, this suggests that for LP 876-10 to be an {\it unbound}
member of the Fomalhaut system in a state of disintegration, then {\it
  we would have to be witnesses to an unusual dynamical state
  predicted to occur over $\sim$0.01-0.1\%\, over the lifetime of
  Fomalhaut A}.  This seems rather unlikely, and the simplest
explanation for the agreement in velocities at the
kilometer-per-second level between LP 876-10 and Fomalhaut A \& B, and
its position within the tidal radius of Fomalhaut AB, is that LP
876-10 is a third bound component of the Fomalhaut system.

\section{Summary}

LP 876-10 is an active (\loglxlbol\, $\simeq$ -3.4), fast-rotating (P
$\simeq$ 0.47 day) star lying within 1 pc of Fomalhaut and TW PsA
(Fomalhaut B), and sharing their motion within $\sim$1 \kms.
\citet{Mamajek12} showed that the isochronal age of Fomalhaut, and
various age diagnostics for TW PsA (rotation, X-ray emission, Li
abundance) were consistent with an age of 440\,$\pm$\,40 Myr for the
pair.  The appearance of LP 876-10 on the main sequence hints that it
is $>$300 Myr in age, and its photometric metallicity ([Fe/H] $\simeq$
-0.1) is in good agreement with spectroscopic metallicity estimates
for TW PsA. We argue that the purported membership of the Fomalhaut
system to the Castor Moving Group does not provide a useful
age-constraint on the system.\\

Based on its position, velocity, and color-magnitude data, we argue
that LP 876-10 is a third stellar component in Fomalhaut system.  The
chances of an interloper field M dwarf sharing the velocity of
Fomalhaut within 1 \kms\, and lying within 1 pc of Fomalhaut is
$<$10$^{-5}$, hence LP 876-10 is almost certainly physically related
to Fomalhaut A and B.  The chances that we are catching the Fomalhaut
system in a state of disintegration, where LP 876-10 is currently
escaping with velocity greater than its predicted escape velocity (0.2
\kms), is statistically unlikely ($<$10$^{-3}$).  Hence, we argue that
LP 876-10 is most likely a {\it bound} low-mass stellar companion to
the Fomalhaut system, which has a well-determined age of
440\,$\pm$\,40 Myr \citep{Mamajek12}. This makes the previously
barely-studied M dwarf LP 876-10 (``Fomalhaut C''), only recently
added to the census of stars within 10 pc via the RECONS astrometry
program, one of the few red dwarfs in the solar neighborhood with a
strongly constrained ($\sim$10\%) age and metallicity ([Fe/H] $\simeq$
-0.1).  Given the difficulty in calibrating the age and metallicity
scale for M dwarfs, Fomalhaut C provides a useful anchor among the
mid-M stars, and another rare of example of a low-mass companion with
separation approaching a parsec. The existence of both Fomalhaut C (LP
876-10) and B (TW PsA) should be considered for future dynamical
calculations trying to explain the unusual offset ($\sim$13 AU)
between Fomalhaut A and its debris disk \citep{Kalas05}, and the
eccentric orbit for the planet candidate Fomalhaut Ab \citep{Kalas13}.

\acknowledgements

We thank the referee for a prompt and thoughtful review which
significantly improved the paper.  We thank Brian Mason, Mark Pecaut,
Valeri Makarov, Massimo Marengo, John Bangert, Christine Hackman,
Demetrios Matsakis, Sean Urban, Alice Quillen, and Paul Kalas for
discussions on LP 876-10, Fomalhaut, and comments on the paper.  EEM
acknowledges support from NSF award AST-1008908.  JLB acknowledges
support from the University of Virginia, Hampden-Sydney College, and
the Levinson Fund of the Peninsula Community Foundation. The RECONS
effort is supported primarily by the National Science Foundation
through grants AST 05-07711 and AST 09-08402. Observations were
initially made possible by NOAO's Survey Program and have continued
via the SMARTS Consortium. This research has made use of NASA ADS,
SIMBAD, Vizier, and data products from 2MASS, WISE, UCAC, SuperWASP,
ASAS, and Hipparcos. This research was made possible through the use
of the AAVSO Photometric All-Sky Survey (APASS), funded by the Robert
Martin Ayers Sciences Fund. This research has made use of the
Washington Double Star Catalog maintained at the U.S. Naval
Observatory.


\begin{thebibliography}{22}
\expandafter\ifx\csname natexlab\endcsname\relax\def\natexlab#1{#1}\fi

\bibitem[Allard et al.(2012)]{Allard12} Allard, F., Homeier, D.,
  Freytag, B., \& Sharp, C.~M.\ 2012, EAS Publications Series, 57, 3

\bibitem[Allen(1963)]{Allen63} Allen, R.~H.\ 1963, Star Names: Their
  Lore and Meaning. New York: Dover, 1963

\bibitem[Baraffe et al.(1998)]{Baraffe98} Baraffe, I., Chabrier, G.,
  Allard, F., \& Hauschildt, P.~H.\ 1998, \aap, 337, 403

\bibitem[Barbier-Brossat \& Figon(2000)]{BarbierBrossat00}
  Barbier-Brossat, M., \& Figon, P.\ 2000, \aaps, 142, 217

\bibitem[Barenfeld et al.(2013)]{Barenfeld13} Barenfeld, S.~A., 
Bubar, E.~J., Mamajek, E.~E., \& Young, P.~A.\ 2013, \apj, 766, 6 

\bibitem[{{Barrado y Navascues}(1998)}]{Barrado98}
{Barrado y Navascues}, D. 1998, \aap, 339, 831

\bibitem[{{Barrado y Navascues} {et~al.}(1997){Barrado y Navascues},
  {Stauffer}, {Hartmann}, \& {Balachandran}}]{Barrado97}
{Barrado y Navascues}, D., {Stauffer}, J.~R., {Hartmann}, L., \&
  {Balachandran}, S.~C. 1997, \apj, 475, 313

\bibitem[Barrado y Navascu{\'e}s et al.(2004)]{Barrado04} Barrado y
  Navascu{\'e}s, D., Stauffer, J.~R., \& Jayawardhana, R.\ 2004, \apj,
  614, 386

\bibitem[Bartlett(2007)]{Bartlett07PhD}
{Bartlett}, J.~L. 2007, PhD thesis, University of Virginia

\bibitem[{{Bartlett} {et~al.}(2007){Bartlett}, {Ianna}, {Henry},
    {Begam}, {Jao}, {Subasavage}, \& {RECONS}}]{Bartlett07AAS}
  {Bartlett}, J.~L., {Ianna}, P.~A., {Henry}, T.~J., et al.\ 2007,
  BAAS, 39, 772

\bibitem[Bean et al.(2010)]{Bean10} Bean, J.~L., Seifahrt, A.,
  Hartman, H., et al.\ 2010, \apj, 713, 410

\bibitem[Bell et al.(2012)]{Bell12} Bell, C.~P.~M., Naylor, T., Mayne,
  N.~J., Jeffries, R.~D., \& Littlefair, S.~P.\ 2012, \mnras, 424,
  3178

\bibitem[Bertin 
\& Arnouts(1996)]{Bertin96} Bertin, E., \& Arnouts, S.\ 1996, \aaps, 117, 393 

\bibitem[Brott \& Hauschildt(2005)]{Brott05} Brott, I., \& Hauschildt,
  P.~H.\ 2005, The Three-Dimensional Universe with Gaia, 576, 565

\bibitem[Burnham(1906)]{Burnham06} Burnham, S.~W.\ 1906, A General
  Catalogue of Double Stars Within 121$^{\circ}$ of the North
  Pole. The Carnegie Institution of Washington.

\bibitem[Burnham(1978)]{Burnham78} Burnham, R., Jr.\ 1978, Burnham's
  Celestial Handbook, Vols.\_1,\_2,\_and\_3., by Burnham, R., Jr..~
  New York (NY, USA): Dover Publ., Inc, 138 p.

\bibitem[{{Busko} \& {Torres}(1978)}]{Busko78}
{Busko}, I.~C. \& {Torres}, C.~A.~O. 1978, \aap, 64, 153

\bibitem[Butters et al.(2010)]{Butters10} Butters, O.~W., West, R.~G.,
  Anderson, D.~R., et al.\ 2010, \aap, 520, L10

\bibitem[Carpenter(2001)]{Carpenter01} Carpenter, J.~M.\ 2001, \aj,
  121, 2851


\bibitem[{{Casagrande} {et~al.}(2011){Casagrande}, {Sch{\"o}nrich},
    {Asplund}, {Cassisi}, {Ram{\'{\i}}rez}, {Mel{\'e}ndez}, {Bensby},
    \& {Feltzing}}]{Casagrande11} {Casagrande}, L., {Sch{\"o}nrich},
  R., {Asplund}, M., et al.\ 2011, \aap, 530, A138

\bibitem[{{Davis} {et~al.}(2005){Davis}, {Richichi}, {Ballester},
    {Gitton}, {Glindemann}, {Morel}, {Schoeller}, {Wittkowski}, \&
    {Paresce}}]{Davis05} {Davis}, J., {Richichi}, A., {Ballester}, P.,
  et al.\ 2005, Astronomische Nachrichten, 326, 25


\bibitem[Delfosse et al.(2000)]{Delfosse00} Delfosse, X., Forveille,
  T., S{\'e}gransan, D., et al.\ 2000, \aap, 364, 217

\bibitem[Dieterich et~al.(2013)]{Dieterich13}Dieterich, S., et al.,
  2013, in prep.

\bibitem[Dotter et al.(2008)]{Dotter08} Dotter, A., Chaboyer, 
B., Jevremovi{\'c}, D., et al.\ 2008, \apjs, 178, 89 

\bibitem[Famaey et al.(2005)]{Famaey05} Famaey, B., Jorissen, A.,
  Luri, X., et al.\ 2005, \aap, 430, 165

\bibitem[{{Fleming} {et~al.}(1995){Fleming}, {Schmitt}, \&
    {Giampapa}}]{Fleming95} {Fleming}, T.~A., {Schmitt}, J.~H.~M.~M.,
  \& {Giampapa}, M.~S. 1995, \apj, 450, 401

\bibitem[Gillett(1986)]{Gillett86} Gillett, F.~C.\ 1986, Light on Dark
  Matter, 124, 61

\bibitem[{{Gontcharov}(2006)}]{Gontcharov06}
{Gontcharov}, G.~A. 2006, Astronomy Letters, 32, 759

\bibitem[{{Gray} \& {Garrison}(1989)}]{GrayGarrison89A}
{Gray}, R.~O. \& {Garrison}, R.~F. 1989, \apjs, 70, 623

\bibitem[Hambly et al.(2001)]{Hambly01} Hambly, N.~C., MacGillivray,
  H.~T., Read, M.~A., et al.\ 2001, \mnras, 326, 1279 (SuperCOSMOS)

\bibitem[Hauschildt et al.(1999)]{Hauschildt99} Hauschildt, P.~H.,
  Allard, F., \& Baron, E.\ 1999, \apj, 512, 377

\bibitem[Heintz(1988)]{Heintz88} Heintz, W.~D.\ 1988, \pasp, 
100, 834 

\bibitem[Henden et al.(2012)]{Henden12} Henden, A.~A., Levine, S.~E.,
  Terrell, D., Smith, T.~C., \& Welch, D.\ 2012, Journal of the
  American Association of Variable Star Observers (JAAVSO), 40, 430


\bibitem[Henry et al.(2006)]{Henry06} Henry, T.~J., Jao, W.-C.,
  Subasavage, J.~P., et al.\ 2006, \aj, 132, 2360

\bibitem[Henry et al.(2004)]{Henry04} Henry, T.~J., Subasavage, J.~P.,
  Brown, M.~A., et al.\ 2004, \aj, 128, 2460

\bibitem[H{\o}g et al.(2000)]{Hog00} H{\o}g, E., Fabricius, C.,
  Makarov, V.~V., et al.\ 2000, \aap, 355, L27

\bibitem[Holman \& Wiegert(1999)]{Holman99} Holman, M.~J., \& Wiegert,
  P.~A.\ 1999, \aj, 117, 621

\bibitem[Irwin et al.(2011)]{Irwin11} Irwin, J., Berta, Z.~K., Burke,
  C.~J., et al.\ 2011, \apj, 727, 56

\bibitem[Jao et al.(2005)]{Jao05} Jao, W.-C., Henry, T.~J., 
Subasavage, J.~P., et al.\ 2005, \aj, 129, 1954 

\bibitem[Jiang \& Tremaine(2010)]{Jiang10} Jiang, Y.-F., \& Tremaine,
  S.\ 2010, \mnras, 401, 977

\bibitem[Johnson \& Apps(2009)]{Johnson09} Johnson, J.~A., \& Apps,
  K.\ 2009, \apj, 699, 933

\bibitem[{{Kaeufl} {et~al.}(2004){Kaeufl}, {Ballester}, {Biereichel},
    {Delabre}, {Donaldson}, {Dorn}, {Fedrigo}, {Finger}, {Fischer},
    {Franza}, {Gojak}, {Huster}, {Jung}, {Lizon}, {Mehrgan}, {Meyer},
    {Moorwood}, {Pirard}, {Paufique}, {Pozna}, {Siebenmorgen},
    {Silber}, {Stegmeier}, \& {Wegerer}}]{Kaufl04} {Kaeufl}, H.-U.,
  {Ballester}, P., {Biereichel}, P., et al.\ 2004, in Society of
  Photo-Optical Instrumentation Engineers (SPIE) Conference Series,
  Vol. 5492, Society of Photo-Optical Instrumentation Engineers (SPIE)
  Conference Series, ed. A.~F.~M. {Moorwood} \& M.~{Iye}, 1218--1227

\bibitem[Kalas et al.(2005)]{Kalas05} Kalas, P., Graham, J.~R., \&
  Clampin, M.\ 2005, \nat, 435, 1067

\bibitem[Kalas et al.(2013)]{Kalas13} Kalas, P., Graham, J.~R.,
  Fitzgerald, M.~P., \& Clampin, M.\ 2013, arXiv:1305.2222

\bibitem[Kalas et al.(2008)]{Kalas08} Kalas, P., Graham, J.~R.,
  Chiang, E., et al.\ 2008, Science, 322, 1345

\bibitem[{{Keenan} \& {McNeil}(1989)}]{Keenan89}
{Keenan}, P.~C. \& {McNeil}, R.~C. 1989, \apjs, 71, 245

\bibitem[Kirkpatrick et al.(2012)]{Kirkpatrick12} Kirkpatrick, J.~D.,
  Gelino, C.~R., Cushing, M.~C., et al.\ 2012, \apj, 753, 156

\bibitem[Landolt(1992)]{Landolt92} Landolt, A.~U.\ 1992, \aj, 
104, 340 

\bibitem[Landolt(2007)]{Landolt07} Landolt, A.~U.\ 2007, \aj, 133,
  2502



\bibitem[L{\'e}pine \& Gaidos(2011)]{Lepine11} L{\'e}pine, S., \&
  Gaidos, E.\ 2011, \aj, 142, 138


\bibitem[{{Luyten}(1938)}]{Luyten38}
{Luyten}, W.~J. 1938, \aj, 47, 115

\bibitem[Luyten \& Hughes(1980)]{Luyten80} Luyten, W.~J., \& Hughes,
  H.~S.\ 1980, Proper Motion Survey with the Forty-Eight Inch Schmidt
  Telescope. LV. First Supplement to the NLTT Catalogue, University of
  Minnesota, 55, 1

\bibitem[Marengo et al.(2009)]{Marengo09} Marengo, M., 
Stapelfeldt, K., Werner, M.~W., et al.\ 2009, \apj, 700, 1647 

\bibitem[Madsen et al.(2002)]{Madsen02} Madsen, S., Dravins, D., \&
  Lindegren, L.\ 2002, \aap, 381, 446

\bibitem[Makarov(2012)]{Makarov12} Makarov, V.~V.\ 2012, \mnras, 421,
  L11

\bibitem[Mamajek(2010)]{Mamajek10} Mamajek, E.~E.\ 2010, Bulletin 
of the American Astronomical Society, 42, \#455.05 

\bibitem[{{Mamajek}(2012)}]{Mamajek12}
{Mamajek}, E.~E. 2012, \apjl, 754, L20

\bibitem[Mart{\'{\i}}n et al.(2006)]{Martin06} Mart{\'{\i}}n, E.~L.,
  Guenther, E., Zapatero Osorio, M.~R., Bouy, H., \& Wainscoat,
  R.\ 2006, \apjl, 644, L75

\bibitem[Mason et al.(2001)]{Mason01} Mason, B.~D., Wycoff, G.~L.,
  Hartkopf, W.~I., Douglass, G.~G., \& Worley, C.~E.\ 2001, \aj, 122,
  3466

\bibitem[{{Mermilliod} \& {Mermilliod}(1994)}]{Mermilliod94}
{Mermilliod}, J.-C. \& {Mermilliod}, M. 1994, {Catalogue of Mean UBV Data on
  Stars}

\bibitem[Monet et al.(2003)]{Monet03} Monet, D.~G., Levine, S.~E.,
  Canzian, B., et al.\ 2003, \aj, 125, 984 (USNO-B1.0)

\bibitem[Monnier et al.(2012)]{Monnier12} Monnier, J.~D., Che, X.,
  Zhao, M., et al.\ 2012, \apjl, 761, L3

\bibitem[Morell(1994)]{Morell94} Morell, O.\ 1994, Ph.D.~Thesis, Acta
  Universitas Upsaliensis

\bibitem[Murgas et al.(2013)]{Murgas13} Murgas, F., Jenkins, J.~S.,
  Rojo, P., Jones, H.~R.~A., \& Pinfield, D.~J.\ 2013, \aap, 552, A27

\bibitem[{{Neuhaeuser} {et~al.}(1995){Neuhaeuser}, {Sterzik},
    {Schmitt}, {Wichmann}, \& {Krautter}}]{Neuhauser95} {Neuhaeuser},
  R., {Sterzik}, M.~F., {Schmitt}, J.~H.~M.~M., et al.\ 1995, \aap,
  297, 391

\bibitem[{{Nordstr{\"o}m} {et~al.}(2004){Nordstr{\"o}m}, {Mayor},
    {Andersen}, {Holmberg}, {Pont}, {J{\o}rgensen}, {Olsen}, {Udry},
    \& {Mowlavi}}]{Nordstrom04} {Nordstr{\"o}m}, B., {Mayor}, M.,
  {Andersen}, J., et al.\ 2004, \aap, 418, 989

\bibitem[Parthasarathy \& Lambert(1987)]{Parthasarathy87}
  Parthasarathy, M., \& Lambert, D.~L.\ 1987, Journal of Astrophysics
  and Astronomy, 8, 51

\bibitem[Pecaut \& Mamajek(2013)]{Pecaut13} Pecaut, M.~J., \& Mamajek,
  E.~E.\ 2013, arXiv:1307.2657

\bibitem[Pojmanski(1997)]{Pojmanski97} Pojmanski, G.\ 1997, Acta
  Astronomica, 47, 467

\bibitem[Pollacco et al.(2006)]{Pollacco06} Pollacco, D.~L., Skillen,
  I., Collier Cameron, A., et al.\ 2006, \pasp, 118, 1407

\bibitem[Press et al.(1992)]{Press92} Press, W.~H., Teukolsky, S.~A.,
  Vetterling, W.~T., \& Flannery, B.~P., Numerical Recipes in C: The
  Art of Scientific Computing, 2nd ed. (New York: Cambridge University
  Press, 1992)

\bibitem[Quillen(2006)]{Quillen06} Quillen, A.~C.\ 2006, \mnras, 372,
  L14

\bibitem[Rajpurohit et al.(2013)]{Rajpurohit13} Rajpurohit, A.~S.,
  Reyl{\'e}, C., Allard, F., et al.\ 2013, \aap, 556, A15

\bibitem[Reid et al.(2002)]{Reid02} Reid, I.~N., Gizis, J.~E., \&
  Hawley, S.~L.\ 2002, \aj, 124, 2721

\bibitem[{{Reid} {et~al.}(2003){Reid}, {Cruz}, {Allen}, {Mungall},
    {Kilkenny}, {Liebert}, {Hawley}, {Fraser}, {Covey}, \&
    {Lowrance}}]{Reid03} {Reid}, I.~N., {Cruz}, K.~L., {Allen}, P., et
  al.\ 2003, \aj, 126, 3007


\bibitem[Ribas(2003)]{Ribas03} Ribas, I.\ 2003, \aap, 400, 297

\bibitem[Roeser et al.(2010)]{Roeser10} Roeser, S., Demleitner, 
M., \& Schilbach, E.\ 2010, \aj, 139, 2440 (PPMXL)

\bibitem[R{\"o}ser et al.(2008)]{Roeser08} R{\"o}ser, S., Schilbach,
  E., Schwan, H., et al.\ 2008, \aap, 488, 401 (PPMX)


\bibitem[Salim \& Gould(2003)]{Salim03} Salim, S., \& Gould, A.\ 2003,
  \apj, 582, 1011

\bibitem[{{Santos} {et~al.}(2004){Santos}, {Israelian}, \& {Mayor}}]{Santos04}
{Santos}, N.~C., {Israelian}, G., \& {Mayor}, M. 2004, \aap, 415, 1153

\bibitem[Schlaufman \& Laughlin(2010)]{Schlaufman10} Schlaufman,
  K.~C., \& Laughlin, G.\ 2010, \aap, 519, A105

\bibitem[Sch{\"o}nrich et al.(2010)]{Schonrich10} Sch{\"o}nrich, R.,
  Binney, J., \& Dehnen, W.\ 2010, \mnras, 403, 1829

\bibitem[Scholz et al.(2005)]{Scholz05} Scholz, R.-D., Meusinger, H.,
  \& Jahrei{\ss}, H.\ 2005, \aap, 442, 211

\bibitem[See(1898)]{See1898} See, T.~J.~J.\ 1898, \aj, 18, 181 

\bibitem[Shaya \& Olling(2011)]{Shaya11} Shaya, E.~J., \& Olling,
  R.~P.\ 2011, \apjs, 192, 2

\bibitem[Skrutskie et al.(2006)]{Skrutskie06} Skrutskie, M.~F., 
Cutri, R.~M., Stiening, R., et al.\ 2006, \aj, 131, 1163 

\bibitem[Soderblom(2010)]{Soderblom10} Soderblom, D.~R.\ 2010, \araa, 48, 581 

\bibitem[Soderblom et al.(2005)]{Soderblom05} Soderblom,
  D.~R., Nelan, E., Benedict, G.~F., et al.\ 2005, \aj, 129, 1616

\bibitem[Stauffer et al.(1998)]{Stauffer98} Stauffer, J.~R., Schultz,
  G., \& Kirkpatrick, J.~D.\ 1998, \apjl, 499, L199

\bibitem[Stauffer et al.(2007)]{Stauffer07} Stauffer, J.~R., Hartmann,
  L.~W., Fazio, G.~G., et al.\ 2007, \apjs, 172, 663

\bibitem[Tinney(1996)]{Tinney96} Tinney, C.~G.\ 1996, \mnras, 
281, 644 

\bibitem[Tokovinin(1998)]{Tokovinin98} Tokovinin, A.~A.\ 1998,
  Astronomy Letters, 24, 178

\bibitem[Townsend et al.(2004)]{Townsend04} Townsend, R.~H.~D., 
Owocki, S.~P., \& Howarth, I.~D.\ 2004, \mnras, 350, 189 

\bibitem[Torres \& Ribas(2002)]{Torres02} Torres, G., \& Ribas,
  I.\ 2002, \apj, 567, 1140


\bibitem[{{van Leeuwen}(2007)}]{vanLeeuwen07}
{van Leeuwen}, F., ed. 2007, Astrophysics and Space Science Library, Vol. 350,
  {Hipparcos, the New Reduction of the Raw Data}

\bibitem[Ventura et al.(1998)]{Ventura98} Ventura, P., Zeppieri, A.,
  Mazzitelli, I., \& D'Antona, F.\ 1998, \aap, 334, 953

\bibitem[{{Voges} {et~al.}(1999){Voges}, {Aschenbach}, {Boller},
  {Br{\"a}uninger}, {Briel}, {Burkert}, {Dennerl}, {Englhauser}, {Gruber},
  {Haberl}, {Hartner}, {Hasinger}, {K{\"u}rster}, {Pfeffermann}, {Pietsch},
  {Predehl}, {Rosso}, {Schmitt}, {Tr{\"u}mper}, \& {Zimmermann}}]{Voges99}
{Voges}, W., {Aschenbach}, B., {Boller}, T., et al.\ 1999, \aap, 349, 389

\bibitem[Winters et al.(2011)]{Winters11} Winters, J.~G., Henry,
  T.~J., Jao, W.-C., et al.\ 2011, \aj, 141, 21

\bibitem[Wright et al.(2010)]{Wright10} Wright, E.~L., Eisenhardt,
  P.~R.~M., Mainzer, A.~K., et al.\ 2010, \aj, 140, 1868

\bibitem[Wroblewski \& Costa(1999)]{Wroblewski99} Wroblewski, H., \&
  Costa, E.\ 1999, \aaps, 139, 25

\bibitem[Yoon et al.(2010)]{Yoon10} Yoon, J., Peterson, D.~M., 
Kurucz, R.~L., \& Zagarello, R.~J.\ 2010, \apj, 708, 71 

\bibitem[Zacharias et al.(2013)]{Zacharias13} Zacharias, N., Finch,
  C.~T., Girard, T.~M., et al.\ 2013, \aj, 145, 44 (UCAC4)

\end{thebibliography}
\end{document}